\pdfoutput=1

\documentclass[
twocolumn,
 amsmath,amssymb,
 aps,
]{revtex4-2}

\usepackage{graphicx}
\usepackage{dcolumn}
\usepackage{bm}

\begin{document}

\title{Integrating two-photon nonlinear spectroscopy of rubidium atoms with silicon photonics}

\author{Artur Skljarow$^{1,2}$,
        Nico Gruhler$^3$, 
        Wolfram Pernice$^3$,
        Harald K\"ubler$^{1,2}$,
        Tilman Pfau$^{1,2}$,
        Robert L\"ow$^{1,2}$,
        and
        Hadiseh Alaeian$^{1,2}$
        }
\affiliation{$^{1}$ 5th Institute of Physics, University of Stuttgart, Pfaffenwaldring 57, 70569 Stuttgart, Germany\\
$ ^{2}$ Center of Integrated Quantum Science and Technology (IQST), Pfaffenwaldring 57, 70569 Stuttgart, Germany\\
$^{3}$ Institute of Physics, University of M\"unster, Hei{\ss}enbergstra{\ss}e 11, 48149 M\"unster, Germany}

\date{\today}

\begin{abstract}
We study an integrated silicon photonic chip, composed of several sub-wavelength ridge waveguides, and immersed in a micro-cell with rubidium vapor. Employing two-photon excitation, including a telecom wavelength, we observe that the waveguide transmission spectrum gets modified when the photonic mode is coupled to rubidium atoms through its evanescent tail. 
Due to the enhanced electric field in the waveguide cladding, the atomic transition can be saturated at a photon number $\approx80$ times less than a free-propagating beam case. The non-linearity of the atom-clad Si-waveguide is about 4 orders of magnitude larger than maximum achievable value in doped Si photonics. The measured spectra corroborate well with a generalized effective susceptibility model that includes the Casimir-Polder potentials, due to the dielectric surface, and the transient interaction between flying atoms and the evanescent waveguide mode. This work paves the way towards a miniaturized, low-power, and integrated hybrid atomic-photonic system compatible with CMOS technologies.
\end{abstract}

\maketitle

\section{Introduction}
The ability to control and exploit light-matter interactions holds substantial promises for discovering novel phenomena and technological applications. From the development of quantum theory, to the state-of-the-art photonic sensors and circuits, novel kinds of light-matter interactions have always opened the door to new fundamental insights and technologies, with interdisciplinary implications reaching beyond physics.

Within the last two decades the noticeable advancements in the field of nano-photonics have broadened the horizon of quantum technologies for promising out-of-the-lab applications~\cite{Politi2008,O'brien2009,Politi2009,Dai2012,Rickman2014,Lin2017}. Integrated photonics allow one to miniaturize and incorporate coherent and non-classical photon sources~\cite{Liang2009,Silverstone2014,Zhou2015}, optically active media~\cite{Liu2009,Yin2012,Wang2013,Isshiki2014}, and detectors~\cite{Vivien2012,Zhang2017} on a single chip. 
Atomic vapors on the other hand, offer a highly polarizable and non-linear medium.
Therefore, combining nano-photonics and atomic physics has promising potentials for devising novel atom-light interaction schemes as well as new quantum hybrid systems benefiting from the best of both worlds.

In recent years, several phenomena such as chiral transport of the photons and their interactions have been studied in a hybrid system of cold atoms interfaced with optical fibers~\cite{Vetsch2010,Goban2012,Mitsch2014, Kato2015,Corzo2016,Kien2017}, and photonic crystal waveguides and cavities~\cite{Thompson2013,Tiecke2014,Goban2015,Perez2017}. Although progress on the miniaturization of cold atom experiments has been reported~\cite{Kulas2017}, combining them in cavity networks is still a challenge.

When combined with nano-photonic circuits, thermal vapors on the other, lend themselves to better scalability and integration~\cite{Uriel2,Uriel3}. So far, thermal atoms have been coupled to integrated waveguides~\cite{Uriel4,Uriel1}, Mach Zehnder interferometer~\cite{Yang2007,Ralf1}, hollow-core fibers~\cite{Slepkov2010,Epple2014}, tapered nano-fibers~\cite{Spillane2008,Hendrickson2010,Fernandez2011}, and ring resonators~\cite{Ralf2}. A recent attempt to increase the coupling strength between atoms and photons benefits from tightly confined mode of a sub-wavelength slot waveguide~\cite{Ralf3}. In spite of less precision and control in thermal vapor cells, their lower technical complexities and better compatibility with integration, still make thermal atoms interesting candidates for realizing non-classical light sources~\cite{Ripka2018}, small-scale atomic clocks~\cite{Newman2019}, and quantum memories~\cite{Guo2019} as some of the major building blocks of a hybrid atom-nanophotonic network.

In this letter, we investigate the interaction between a thermal rubidium (Rb) vapor and sub-wavelength ridge waveguides. To benefit from the well-established CMOS technology in silicon (Si) and be compatible with the telecommunication band, we use a two-photon excitation scheme, i.e. ($5S_{1/2} \rightarrow 5P_{3/2}$) at $\lambda_1$ = 780 nm and ($5P_{3/2} \rightarrow 4D_{5/2}$) at $\lambda_2$ = 1529 nm (telecom wavelength) where the latter is coupled to the wave\-guide. We demonstrate the coupling of rubidium atoms to the evanescent tail of the waveguide mode in the cladding. By systematic measurement of the absorption linewidth at different telecom laser powers, even down to the single photon level, we investigated the effect of tight-mode confinement of the sub-wavelength waveguide via a noticeable reduction of the saturation power. In addition, we investigate the effect of the surface on the atomic spectra via the Casimir-Polder (CP) potential. Our experimental measurements can be properly described via a generalized effective susceptibility model based on total internal reflection (TIR)-spectroscopy~\cite{Ducloy} when CP potentials and transient effects due to the atomic motions are incorporated.

Although the coupling between atoms and the waveguide mode is still in the weak regime  we believe that our scheme has the potential of investigating stronger coupling realms when more confined geometries like slot waveguides and cavities~\cite{alaeian2019cavity} are considered. Moreover, the substantial reduction of the required power for inducing optical non-linearity makes this hybrid device an efficient candidate to produce non-classical light at the telecommunication band~\cite{Moon2019}.

\section{Measurement Scheme}
\begin{figure}[htbp]
\centering
\includegraphics[width=8.21cm]{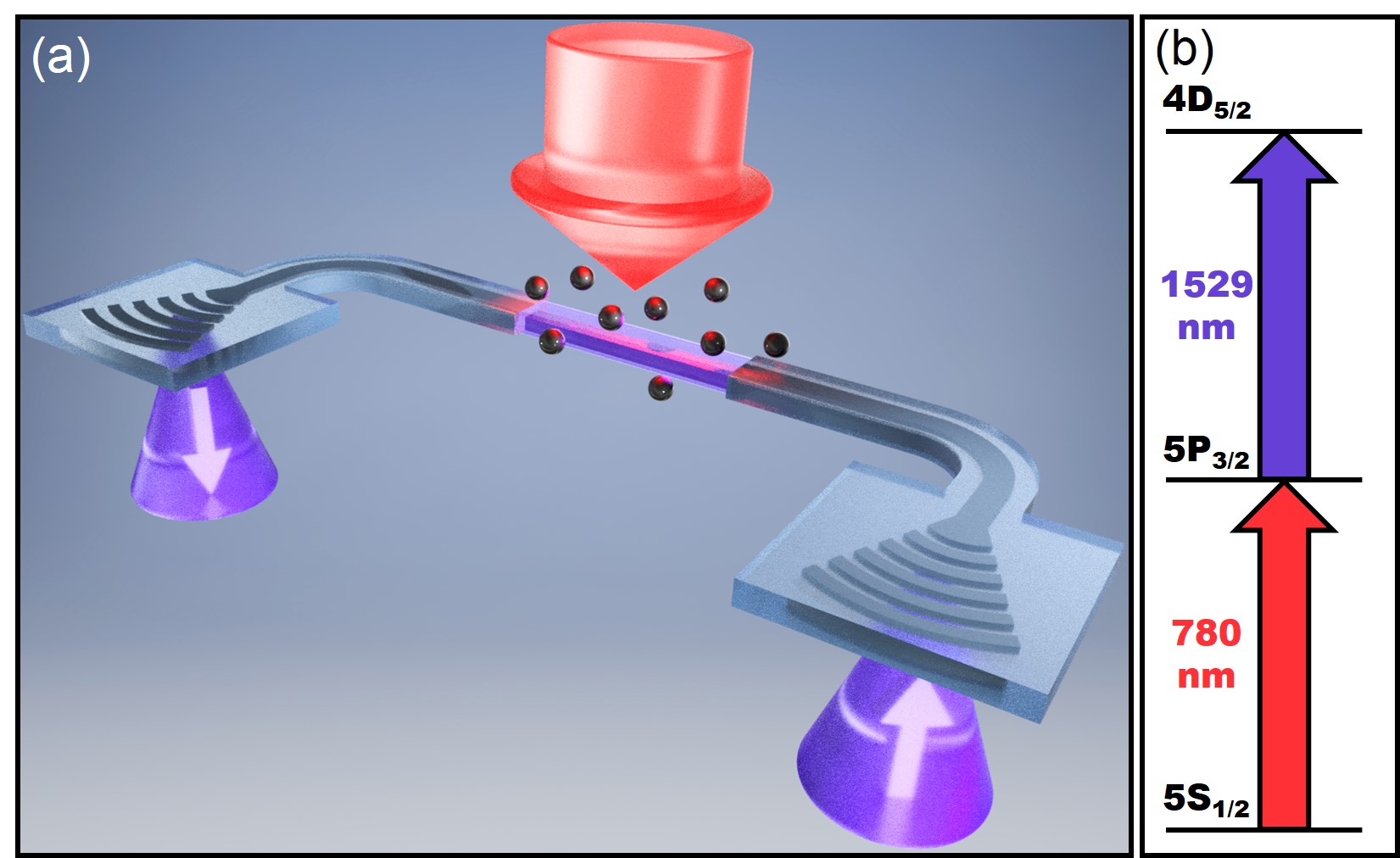}
\caption{ (a) Artistic depiction of the experiment. The 780 pump laser (red) illuminates the nano-device perpendicular to the chip surface from the top and excites Rb atoms to 5P$_{3/2}$ state. The 1529 nm probe laser (purple) is coupled to the silicon waveguide via grating couplers from below. Within the uncovered interaction region the evanescent tail of the guided mode excites the atoms from the intermediate level to the 4D$_{5/2}$ state. (b) Simplified level scheme of the $^{85}$Rb isotope.}  
\label{fig:setup}
\end{figure}

Fig.~\ref{fig:setup}(a) and (b) show the schematics of the experiment and simplified electronic levels of $^{85}$Rb isotope, respectively. A 780 nm pump laser (Toptica DLX110, red), locked to (5S$_{1/2}$ F = 3 $\rightarrow$ 5P$_{3/2}$ F' = 4) transition of $^{85}$Rb isotope, uniformly illuminates the un-covered part of the waveguide and excites Rb atoms to the intermediate excited state. A 1529 nm probe laser (Toptica DL Pro, purple) is coupled to the waveguide via a grating coupler and scanned over 4D$_{5/2}$ states. The evanescent tail of the waveguide mode interacts with excited Rb atoms in the cladding and further excites them to 4D$_{5/2}$ level. The modified mode gets coupled out of the waveguide via another grating coupler and recorded with an avalanche photodiode (APD, THORLABS APD410C) or a single photon detection module (SPCM, ID Quantique id210) at very low powers. A detailed description of the setup and nano-photonic cell can be found in the appendix section A and B.

\section{Numerical simulation and fit of the evanescent waveguide spectrum}
\begin{figure}[htbp]
\centering
\includegraphics[width=\linewidth]{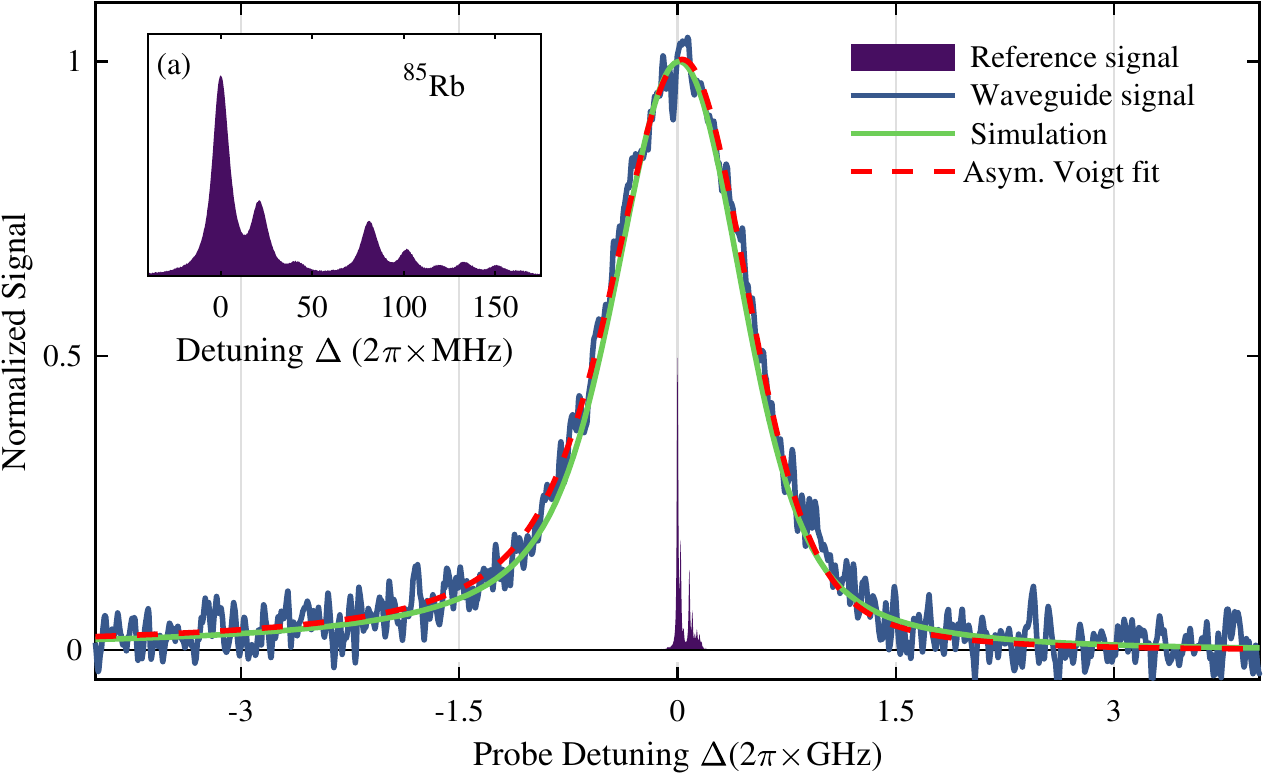}
\caption{ Normalized absorption of the 1529 nm probe laser traveling through a Si waveguide as a function of detuning (blue). The green line shows the calculated effective susceptibility using Eq.(\ref{eq:Xeff}) and the dahsed line shows the asymmetric Voigt fit to the measured data (blue). The hyperfine reference signal (purple area) is from a counter-propagating pump and probe scheme in a 10 cm Rb cell.
Inset (a) shows a zoomed-in version of the reference spectrum with all hyperfine transitions.}
\label{fig:sim}
\end{figure}

The blue trace in Fig.~\ref{fig:sim} shows the measured absorption of the probe laser after lock-in-amplification of the APD signal averaged over 2000 traces at a mode power of 41 fW (below saturation) and a temperature of 130$^\circ$C. 
Inset (a) shows the measured absorption of a free propagating probe beam through a 10 cm long, cylindrical "reference" cell for the $^{85}$Rb isotope with free counter-propagating pump beam. 
In the following, the three main differences between reference and waveguide signal are explained. The first difference arises from the wavevector orientation of pump and probe beam and the loss of Doppler selectivity. In case of the waveguide signal the wavevectors are orthogonal. This means the pump laser addresses all velocity classes.
In addition, the wavevector of the probe laser is modified by the effective refractive index $n_{\text{eff}} = 2.53$ of the waveguide mode which contributes to the Doppler broadening $\omega_{\text{Dopp}} = (n_{\text{eff}}\sigma)/\lambda_2$, with $\sigma$ being the mean velocity of the atoms in the vapor. Together with the wavevector orientation, this leads to different spacing of the individual hyperfine peaks and Doppler broadening of the whole spectrum.

The second difference arises from the short transit time of flying atoms through an evanescent field with a decay length of $L = 1/k_\perp \approx 90$ nm, where $k_\perp$ is the normal component of the wavevector. These atoms experience a broadening of about $\Gamma_{\text{tt}}/(2\pi) = 435$ MHz.

The third difference stems from the Si surface due to the CP potentials which leads to an overall shift and asymmetry of the lineshape.

To obtain a more quantitative understanding we employ an effective susceptibility $\chi_{\text{eff}}$ description to model the effect of the waveguide mode in polarizing Rb atoms. With this theoretical model we assume that a certain amount of the atoms are pumped to the 5P$_{3/2}$ state, with all m$_\text{F}$ levels equally populated, and we only consider the 5P to 4D transition, summing over all allowed transitions with weights calculated from Clebsch-Gordon (CG) coefficients.
Each transition is modeled as a two-level system and in the non-depleted regime which is a valid assumption in the weak probe limit when the optical pumping effects can be safely ignored. This leads to the following expression for the effective susceptibility of the atoms\\
\begin{widetext}
\begin{align}
\label{eq:Xeff}
\chi_{\text{eff}}(\Delta) =  A \int \int d\boldsymbol{v} dl F(\boldsymbol{v},l) \times
&\left[\Theta(v_\perp) \frac{e^{-l k_\perp} - e^{-{l/v_\perp[\Gamma_0/2 + i (\Delta - C_3/l^3 - v_\parallel k_\parallel)]}}}{(\Gamma_0/2 - v_\perp k_\perp) + i (\Delta -C_3/l^3 - v_\parallel k_\parallel)} + \right.\\ \nonumber
&\left. \Theta(-v_\perp) \frac{e^{-l k_\perp}}{(\Gamma_0/2 - v_\perp k_\perp) + i (\Delta - C_3/l^3 - v_\parallel k_\parallel)} \right].
\end{align}
\end{widetext}
In this equation, $l$ is the distance of the atom from the surface.\\
$F = \exp{[-v^2/(2\sigma^2)]}/(2\pi \sigma^2)\exp{(-l k_\perp)}/(2k_\perp)$ contains the Boltzmann factor for velocities perpendicular ($v_\perp$) and along the waveguide surface ($v_\parallel$) and the field decay length normal to the waveguide in y,z-directions (see the inset of Fig.~\ref{fig:sat}(b) showing the profile of the waveguide and the coordinates).

The denominator of Eq. \ref{eq:Xeff} contains the natural decay rate $\Gamma_0/(2\pi) = 1.89$ MHz for the 4D$_{5/2}$ state.
Doppler and transit time effects are included via $v_\parallel k_\parallel$ and $v_\perp k_\perp$ terms, respectively. 
Close to the surface, Rb atoms induce a mirrored dipole with opposite charge. The position dependent (attractive) dipole-dipole interaction is described by the CP potential.
From numerical calculation, we found that a $C_3l^{-3}$ potential is a good estimate to calculate the surface effect on the atomic level shift within the field decay length.
A detailed description of the full CP potential can be found in the appendix section C.

Our simulations show that the cyclic transition (5P$_{3/2}$ F = 4 to 4D$_{5/2}$ F = 5) (largest peak in Fig.~\ref{fig:sim}(a)) dominates the broadened waveguide signal. Each hyperfine transition is red shifted by $-2\pi\times53$~MHz. However, the full spectrum including all other hyperfine transitions shows a shift of $-2\pi\times7$~MHz. This is because the weights of the higher lying transitions shift the overall spectrum by +46 MHz in the opposite direction without significant change of the lineshape. Therefore, the waveguide spectrum in Fig.~\ref{fig:sim} shows no significant shift.

In contrast to atoms moving parallel to the surface and experiencing a uniform field, the perpendicular direction exhibits an exponential field gradient with different initial conditions selected by the Heaviside function $\Theta$;
atoms traveling away from the structure ($v_\perp>0$) have left the surface in the 5P$_{3/2}$ state, while atoms moving towards the surface ($v_\perp<0$) continuously feel an increasing field strength.
The susceptibility is calculated for all nine hyperfine transitions with corresponding amplitudes $A_j$ of the $j^{th}$ transition set via CG-coefficients.
The simulation in Fig.~\ref{fig:sim} shows the full spectrum. 
To extract physical quantities we fit an asymmetric Voigt profile to the measured data \cite{AsymVoigt}. The fit function is a convolution of symmetric Gaussian function and a Lorentzian function whose width (FWHM) is modified by two asymmetry parameters \cite{NumVoigt}.

Since the Gaussian contribution (Doppler effect) as well as the asymmetry parameters (surface effects) are the same for all measurements, the Gaussian width is fixed to $\omega_\text{G} = (2\pi n_{\text{eff}})/\lambda \times \sigma = 2\pi\times784$ MHz, a value corresponding to a cell temperature of 140$^\circ$C.
The asymmetry parameters were empirically determined from the asymmetric Voigt fit for one of the measurements and kept fixed for the rest of the analysis since the geometry hence, the surface effects remain the same (fits to various spectra show consistent values for the asymmetry parameters).

We confirmed with both the simulation and the fit that the cyclic transition (5P$_{3/2}$ F = 4 $\rightarrow$ 4D$_{5/2}$ F' = 5) dominates the absorption lineshape, so one asymmetric Voigt fit is sufficient to fit the spectrum with the Lorentzian width, amplitude, and offset as the only free fit parameters.
From the good agreement, we deduce that the asymmetric Voigt fit offers a proper description to extract information about absorption line-width and non-linearity from our measurements also at high probe power.

\section{Probe Power Effect, Broadening, and Non-Linearity}

\begin{figure}[htbp]
\centering
\includegraphics[width=\linewidth]{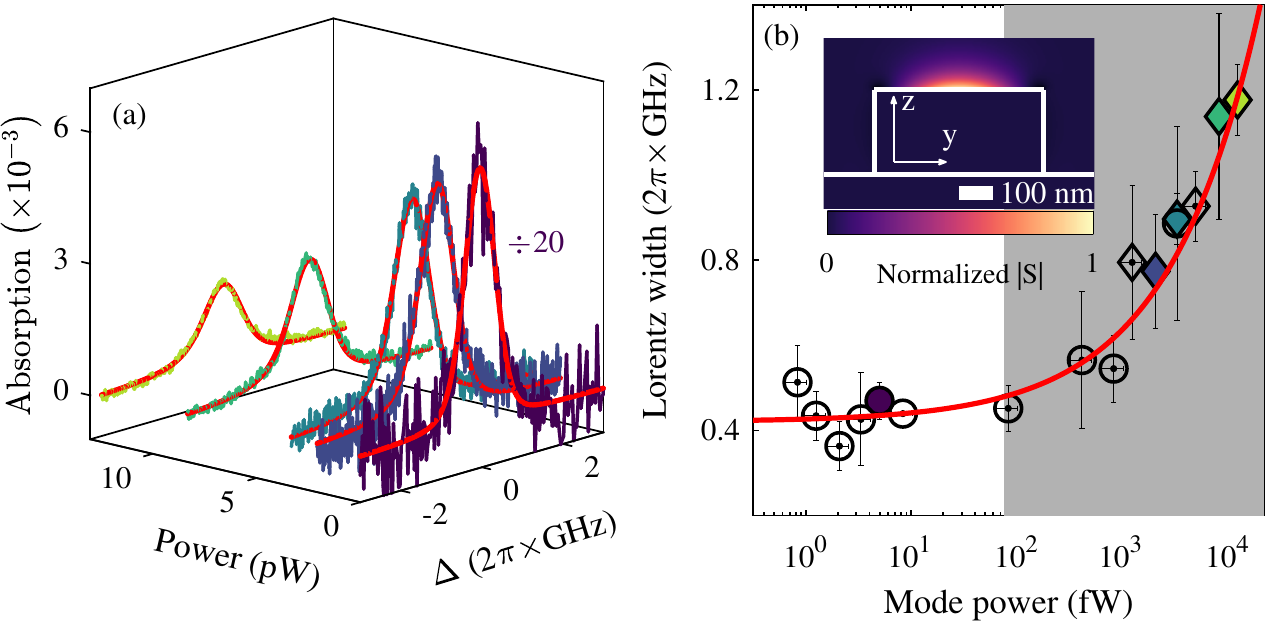}
\caption{ (a) Measured absorption for different probe powers as a function of probe detuning $\Delta$. The purple trace at lowest power is recorded with the SPCM and scaled down by a factor of 20 for comparison to other traces taken with the APD at higher powers. Red lines show asymmetric Voigt fits to the measured data. 
(b) The mean Lorentzian FWHM vs. probe power interacting with rubidium atoms. The vertical errorbars indicate standard deviations from 4 statistical measurements for each power while the horizontal errorbars result from the fit. 
Circles and diamonds stand for SPCM and APD measurements, respectively. 
The filled points are color-coded as the spectra in panel (a).
The solid red line shows the fit function for homogeneous power broadening vs. probe power.
The gray area shows the saturated region. Inset: Waveguide geometry and the real part of the Poynting vector outside the waveguide and along its axis.
}
\label{fig:sat}
\end{figure}

Tight mode confinement within the waveguide leads to much higher intensities compared to free propagating laser beams. In this section we study the non-linearity of the hybrid system and calculate the effective power and photon flux needed for observing this non-linear behaviour.

The inset of Fig.~\ref{fig:sat}(b) shows the COMSOL simulations of the mode Poynting vector ($S$) distribution, carrying the power along the waveguide and within the cladding. This quantity allows one to define an effective mode area interacting with atoms $A_{\text{eff}}$ at 1529 nm as
\begin{equation}
A_{\text{eff}} = \frac{1}{\max(S)}\int_{A_{\text{out}}} SdA = 3510\,\text{nm}^2
\end{equation}
where $\text{max}(S)$ is the global maximum of the Poynting vector and $A_{\text{out}}$ denotes the region outside the waveguide.
Systematic measurements of the absorption spectrum as a function of the probe laser power allows one to determine the saturation power, hence the \emph{real} power interacting with atoms.

Figure~\ref{fig:sat} (a) shows measured spectra for different probe powers at 140$^\circ$C cell temperature and a ground state density of $n_{\text{Rb}} \approx 10^{13} \text{cm}^{-3}$ (A detailed study for different atomic densities and their effect on the pump laser can be found in the appendix section E).
The purple trace, corresponding to the lowest power, is recorded with a single photon counting module. The obtained histogram from the photon counter has been translated directly to an absorption spectrum (more information can be found in the appendix section D). 

All other traces in Fig.~\ref{fig:sat} (a) at higher powers were obtained with an APD.
The spectra were fitted using asymmetric Voigt fit functions with the same asymmetry parameters determined in the previous section. Figure~\ref{fig:sat}(b) shows the fitted Lorentzian FWHM as a function of the probe power. Vertical errorbars are obtained from statistical averages over four measurements with identical experimental parameters.
Diamond-shaped data points correspond to spectra measured with the APD and the circle-shaped data are from the SPCM measurement.
From a series of probe power-dependent measurements in a Rb reference cell we obtained the saturation intensity of $I_{\text{sat}} \approx 0.138$ mW/cm$^2$ for the (5P$_{3/2}$ $\rightarrow$ 4D$_{5/2}$) transition, the value that has been confirmed with calculations based on the transition dipole moments and lifetime, as well.

In each experiment, we first measured the probe power $P_{\text{ref}}$ in the focus of the magnifying telescope (see $P_{\text{ref}}$ in Fig.~\ref{fig:setup} of appendix). Later, the power dependency of the Lorentzian FWHM was fitted with $\Gamma = \Gamma_0\sqrt{1+\gamma P_{\text{ref}}/(A_{\text{eff}}I_{\text{sat}})}+\Gamma_{\text{off}}$ to extract the power broadening behavior. In this expression $\gamma$ is a scaling factor that captures all losses (including optics and coupling to nano-structure) from the point where $P_{\text{ref}}$ is measured to the interaction region. 
The offset $\Gamma_{\text{off}}/(2\pi) = 420$ MHz is close to the value of transit time broadening $\Gamma_{\text{tt}}$ obtained from the effective susceptibility simulation.
The red curve in Fig.~\ref{fig:sat} (b) shows $\Gamma = \Gamma_0\sqrt{1+I/I_{\text{sat}}}+\Gamma_{\text{off}}$ as a function of the mode power $P_{\text{mode}} = \gamma P_{\text{ref}}\times A_{\text{tot}}/A_{\text{eff}}$.
It must be noted that while $A_{\text{tot}}$ is the total effective mode area however, only a fraction of $A_{\text{eff}}/A_{\text{tot}} = 6.3\%$ of the mode power interacts with the surrounding Rb atoms.

The filled data points are color-coded as the spectra depicted in Fig.~\ref{fig:sat} (a).
The obtained APD traces show only relative absorption after the lock-in amplifier. 
In order to obtain the absolute absorption values for a broad range of intensities, below and above the saturation power of $P_{\text{sat}} = 76 \,\text{fW}$  (gray shaded area in Fig.~\ref{fig:sat}(b)), the absorption coefficient $\alpha = \alpha_0/(1+I/I_{\text{sat}})$ was calculated for the filled data points.
Here, $\alpha_0$ is the power independent absorption coefficient in the weak probe regime. 
The SPCM measurement gives the intensity-independent absorption $\alpha_0$ and the APD spectra are scaled according to the calibrated input power with $\alpha(I)$. 
For visualization purposes the SPCM trace is scaled down by a factor of 20 compared to the APD traces.

In order to investigate the onset of optical non-linearity at low photon number achievable in our system we first calculate the $\chi^{(3)}$ coefficient contributing to the atomic polarizability as $P = \epsilon_0[\chi^{(1)}+3/4 \chi^{(3)}E^2]E$. Following the two-level system approach (5P$_{3/2}$ $\rightarrow$ 4D$_{5/2}$) we obtain $|\text{Re}(\chi^{(3)}(\Delta))]|/n_{\text{Rb}} = 9.77\times10^{-32} \text{m}^5/\text{V}^2$ at $|\Delta| = 2\pi\times 1 $GHz. This large detuning has been considered to be far away from any single-photon absorption of the Doppler broadened spectrum of the thermal vapor. This non-linearity corresponds to a Kerr coefficient of $n_2 \approx 3.68\times 10^{-6} \text{cm}^2/\text{W}$, for a vapor at 150$^\circ$C with the density of $n_{\text{Rb}} = 10^{19}$ m$^{-3}$ at intermediate $5P_{3/2}$-state. For comparison, a free-propagating beam focused down to the diffraction limit has a focus spot of $A_{\text{Gauss}} = \pi/2\times(\lambda/2)^2$. This means in the vicinity of the waveguide $A_{\text{Gauss}}/A_{\text{out}} = 78$ times less photons are needed to reach this non-linear regime in contrast to a free propagating beam. As mentioned before, at $P= 76$ fW of the mode power the upper transition gets saturated corresponding to a photon flux of $\Phi = 5.8 \times 10^5 s^{-1}$.
Not only this non-linearity is approximately 7 orders of magnitude higher than the achievable values for bulk silicon~\cite{Kerr3}, the enhanced field density of the waveguide allows one to reach this value at substantially lower powers of fW-range.
An effective way to overcome the weak intrinsic non-linearity of silicon is to integrate it with materials with a high Kerr coefficient.  Using silicon-doped and silicon-graphene hybrid systems the Kerr non-linearity has been  increased  up to $n_2 \approx 2\times 10^{-13} cm^2/W$, about an order of magnitude enhancement compared to bulk silicon~\cite{Feng2019}. In an atom-clad waveguide the non-linear medium is only interacting with a small fraction of the mode power (less than $10\%$ in our case) however, the optical non-linearity remains large enough to exceed those values by several orders of magnitude. Although the gain bandwidth of the atomic medium is limited to a few GHz, atom-photonic hybrid systems are still promising candidates for non-linear photonics.

\section*{Conclusion}

In this letter we investigated a hybrid system consisting of thermal alkali vapor and integrated on-chip silicon photonic structures. Via two-photon spectroscopy of rubidium atoms we investigated the effect of atomic polarizability on modifying the waveguide transmission~\cite{TwoPhotonSRS}. The spectra show an increased Doppler broadening, due to the large effective index of the waveguide mode, accompanied with an additional broadening due to the transit interaction of moving atoms with the mode. In addition, due to the CP potential, the transmission spectra becomes shifted and asymmetric~\cite{Failache2003}. Using an effective susceptibility model, including all relevant atomic transitions, modified Doppler effect, transit effect, and CP potential we have been able to reproduce the measured data with excellent agreement. From probe power measurements beyond saturation we extracted the non-linear response of our system in terms of a $\chi^{(3)}$ coefficient. We demonstrated that large non-linearity at telecom wavelength on a silicon platform can be reached with powers on the order of fW. This non-linearity is 4 orders of magnitude larger than state of the art values achievable in hybrid silicon systems.

Unlike previous reports on Si$_3$N$_4$ photonics~\cite{Ralf1} we did not observe any noticeable degradation of devices within the period of 5-6 months. This could be attributed to the thin native oxide layer (SiO$_2$) on silicon, protecting silicon from rubidium atoms.
This hybrid system opens the door for future experiments with various waveguide geometries with applications for fast optical switching, cross-phase modulation and photonic gate operations. On a more fundamental level, this system is suited to study interacting atoms in a 1D situation, eg. slot waveguides~\cite{Ralf3}.

The small mode area of the evanescent field, on the order of the wavelength, allows for systematic studies of atom-atom interaction effects like self-broadening in low dimensional geometries. 
The extension of the evanescent field can be enhanced by tapering the waveguides facilitating two-photon excitation to the Rydberg levels, hence creating large optical non-linearity. These interactions could also be utilized to add non-linearity, e.g. to a system of coupled ring resonators, which on their own create a synthetic gauge field for non-interacting photons~\cite{Hafezi2013}.

\section{Appendix}
\subsection{Cell Design}
\begin{figure}[htbp]
\centering
\includegraphics[width=8.21cm]{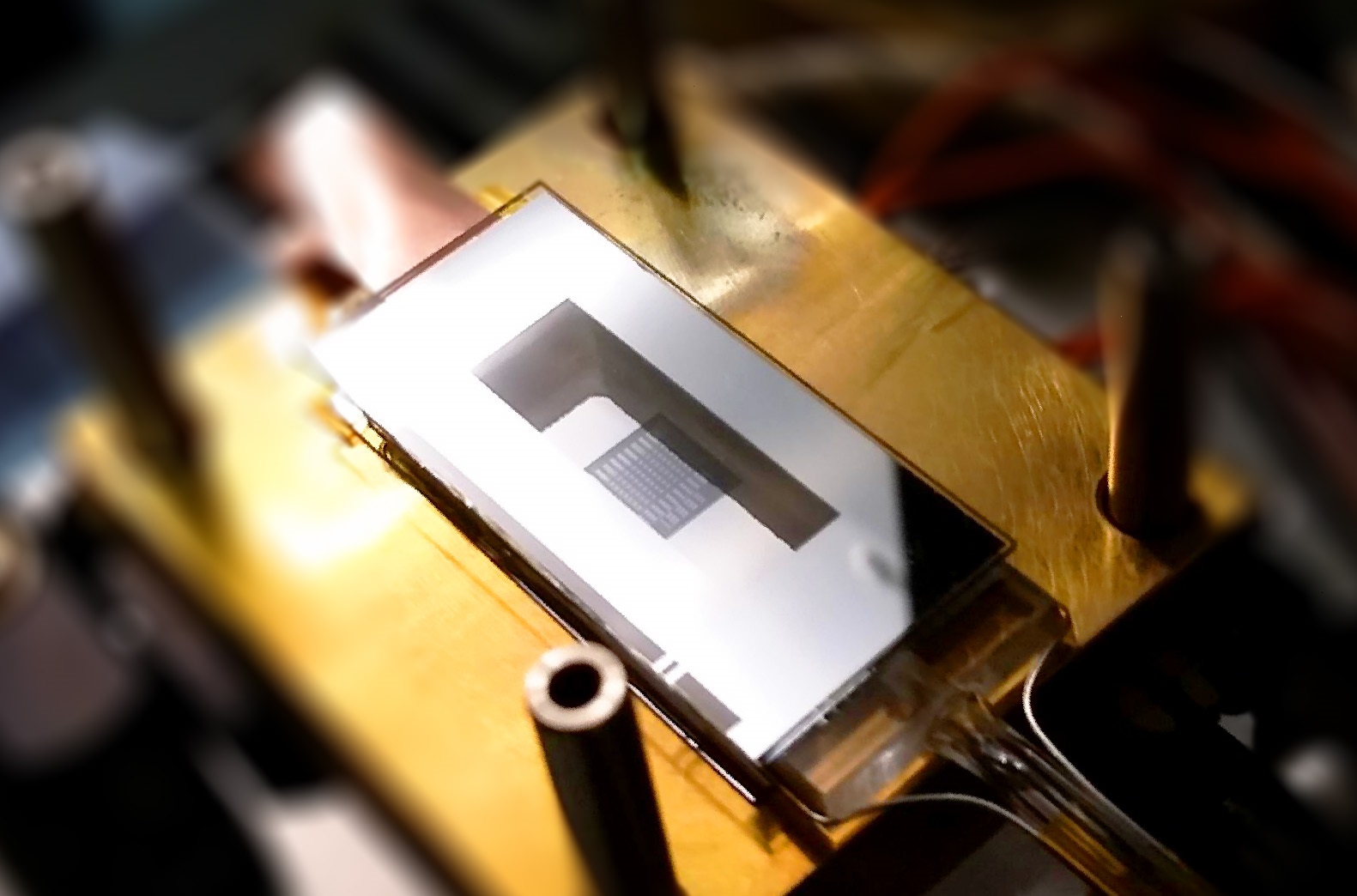}
\caption{Photograph of the home-built Rb vapor cell.}
\label{fig:Cell}
\end{figure}
A photograph of the vapor cell is depicted in Fig.~\ref{fig:Cell}. 
The 1$\times$1 cm nano-photonic chip on top of the Si window is visible through the glass window.
Around the glass window a thin layer of Al is deposited.
This way by applying a negative voltage at 330$^\circ$C to the glass frame and a positive voltage to the windows we force ion diffusion through the cell.
Negatively charged O$^{2-}$ ions from the glass frame form bonding layers at the windows with Si and Al, respectively. The Rb reservoir is added after evacuation.

\subsection{Experimental Setup}
\begin{figure}[htbp]
\centering
\includegraphics[width=\linewidth]{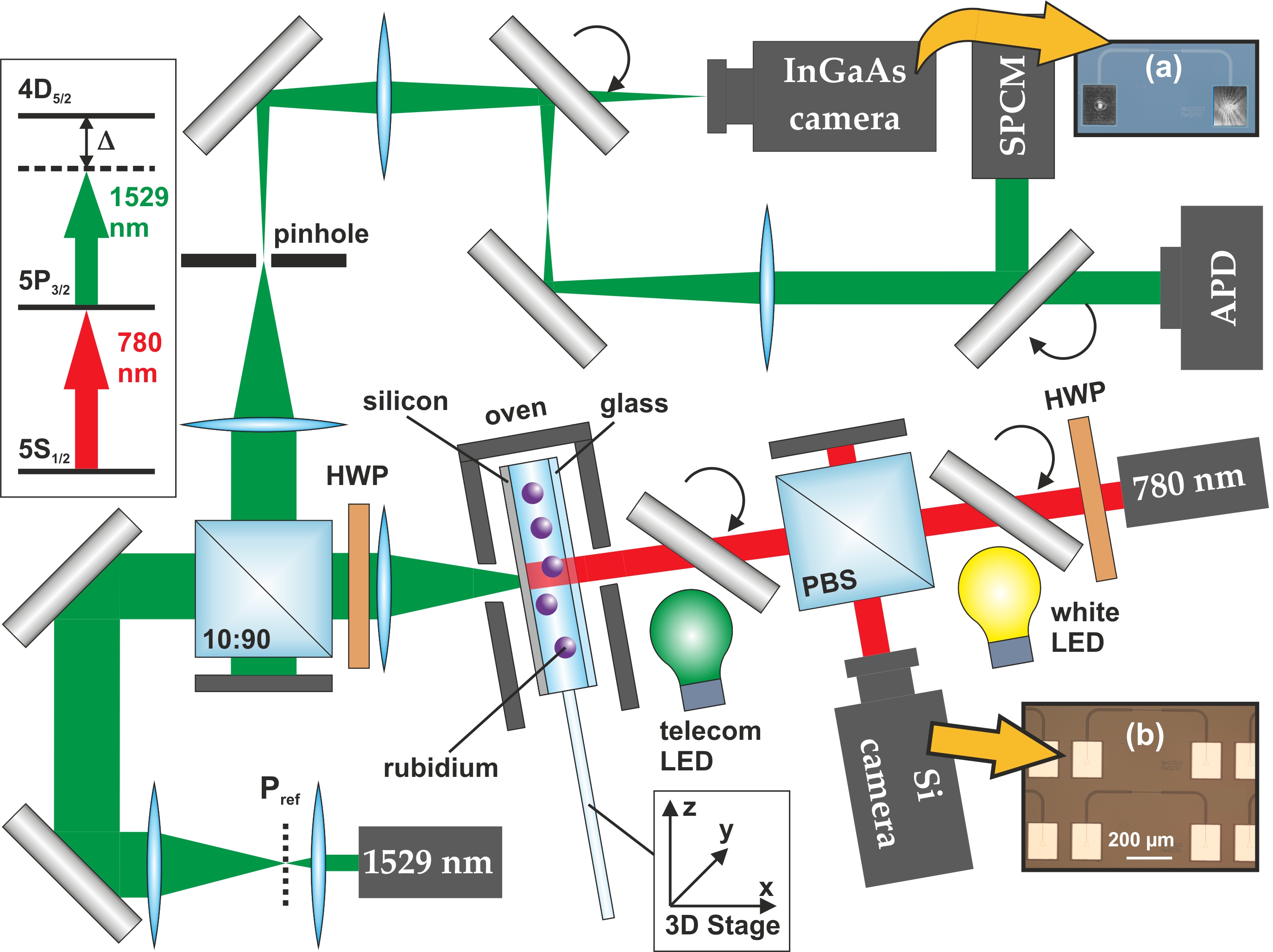}
\caption{Sketch of the excitation scheme, imaging setup, and Rb vapor cell. 
A 3D translational stage allows addressing different devices on the Si chip. 
The 780 nm pump laser illuminates the cell through the glass window and excites atoms from the ground (5S) to the 5P state.
The 1529 nm probe laser is coupled into the nano-device and its evanescent interaction with the atoms in the 5P state is recorded with an APD and SPCM.
(a) Image from the InGaAs camera with light coupled in (left) and out (right). 
(b) The view through the glass window allows alignment of the pump laser.
}
\label{fig:setup_sketch}
\end{figure}

The investigated nano-device is a 1 mm long, $500\times250$ nm rectangular Si waveguide between two grating couplers supporting only one TE00 mode at 1529 nm on a Si0$_2$/Si substrate (see Fig.~\ref{fig:setup_sketch} inset (a)).
This $2\times5$ cm substrate is anodically bonded to a glass frame and acts as a Si window.
The other side of the glass frame is covered by a glass window. 
A glass tube attached to the frame contains the Rb reservoir and is mounted on a 3D translational stage in the center of Fig.~\ref{fig:setup_sketch}.
In order to tune the atomic density and prevent condensation on the Si chip, the reservoir temperature is independently controlled.
The chosen level scheme of $^{85}$Rb is depicted Fig.~\ref{fig:setup_sketch}. A 780 nm pump laser (Toptica DLX110, red) is locked to the 5S$_{1/2}$ F = 3 $\rightarrow$ 5P$_{3/2}$ F' = 4 transition while the 1529 nm probe laser (Toptica DL Pro, green) scans over the 4D$_{5/2}$ states.
Both lasers are operating in continuous wave mode.
The red pump laser is illuminating the whole structure through the glass window of the cell.
This way, the pump power is distributed homogeneously across the structure.
A Si camera monitors the pump spot on the chip (see inset (b)). 
The probe laser is coupled into the nano-device through 10$^\circ$ tilted on-chip grating couplers. The halfwave plate matches the polarisation to the guided mode.
Light coming from the out-coupler is 90$\%$ reflected at the 10:90 beamsplitter and spatially filtered by the pinhole before being imaged on the InGaAs camera (Xenics Bobcat 320 Gated).
Evanescent two-photon interaction of the guided light with the atoms is recorded with an avalanche photodiode (APD, THORLABS APD410C) and at lower light level with a single photon detection module (SPCM, ID Quantique id210). 
When measured with an APD below the breakdown voltage, the pump laser is modulated for lock-in amplification using an AOM in combination with a lock-in amplifier (Stanford Research Systems SR830).
The SPCM is used together with a time tagger (Swabian Instruments TT20) to produce a histogram for each laser scan.
The LEDs act as background illumination for alignment issues.
Frequency axes are calibrated with a Fabry-P\'erot interferometer with a free spectral range of 2030 MHz.

\subsection{Casimir Polder Potential}
\begin{figure}[htbp]
\includegraphics[width=\columnwidth]{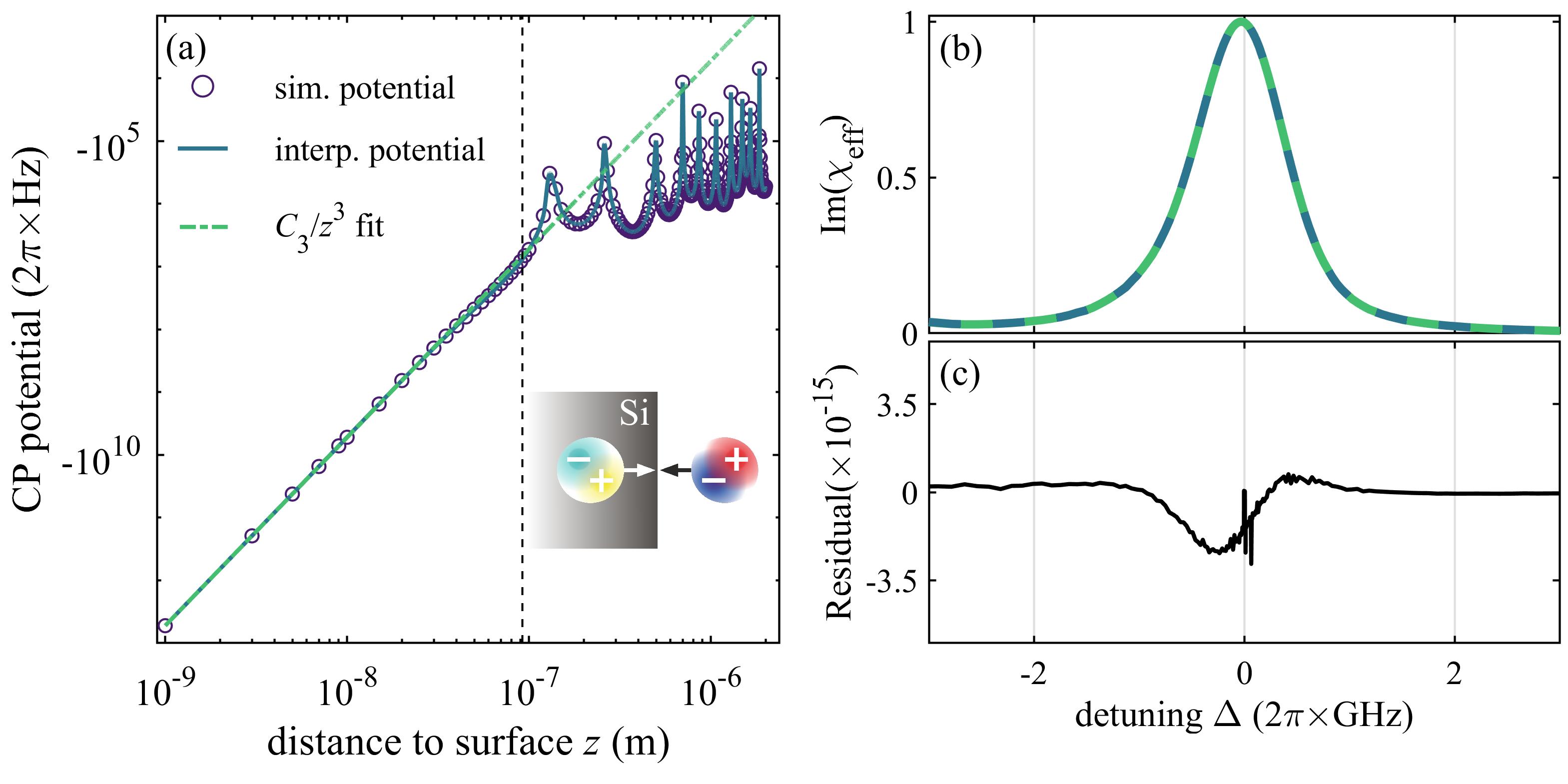}
\caption{(a) The simulated potential using the local density of states (LDOS) approach is depicted as circles. 
This potential is then interpolated (blue line) and approximated with a C$_3$ fit (green dashed line).
The black dashed line marks the decay length of the evanescent field.
The inset illustrates the cause of the potential. 
(b) Effective susceptibility plotted as function of detuning using the interpolated potential and the $C_3$ approach. Colors match with the legend in (a). 
(c) Difference in effective susceptibility between the interpolated and $C_3$ fit approach as a function of detuning.}
\label{fig:CP}
\end{figure}

The origin of the Casimir-Polder (CP) effect is sketched in the inset of Fig.~\ref{fig:CP}(a). 
An atomic dipole sees its mirrored dipole in the vicinity of a dielectric surface.
This leads to an interaction potential between the dipole and its induced image.
However, this effect is only noticeable on a short distance away from the surface. 
Fig.~\ref{fig:CP}(a) shows the induced CP lineshift (circles) on a Rb atom in dependence of its distance $z$ to the Si surface for the (5P$_{3/2}$ $\rightarrow$ 4D$_{5/2}$) transition at 1529 nm.
Up to 100 nm the potential is proportional to $z^{-3}$. 
At longer distances, retardation effects lead to deviations from the $z^{-3}$-behavior.
In order to investigate the influence of the retardation, we interpolated the potential (blue line) and fitted the potential up to 100 nm with $f(z) = C_3/z^3$ (green dashed line) and included these values in the numerical calculation of the susceptibility.
The results are shown in Fig.~\ref{fig:CP}(b) with colors matching the legend in (a). 
For a fair comparison the $z$ integration was performed from 0 to 2 $\mu$m.
Figure~\ref{fig:CP}(c) shows that both approaches give identical results.
Contributions to $\chi_{\text{eff}}$ from atoms at a larger distance ($z>1\mu$m) from the surface are comparably small and additionally attenuated by the evanescent decay of the mode field (black dashed line in Fig.~\ref{fig:CP}(a)).
Hence, the $C_3/z^3$ potential used in the main text is a valid approximation to treat the surface effect properly.

\subsection{Generating Spectra with the SPCM}

\begin{figure}[htbp]
\centering
\includegraphics[width=8.21cm]{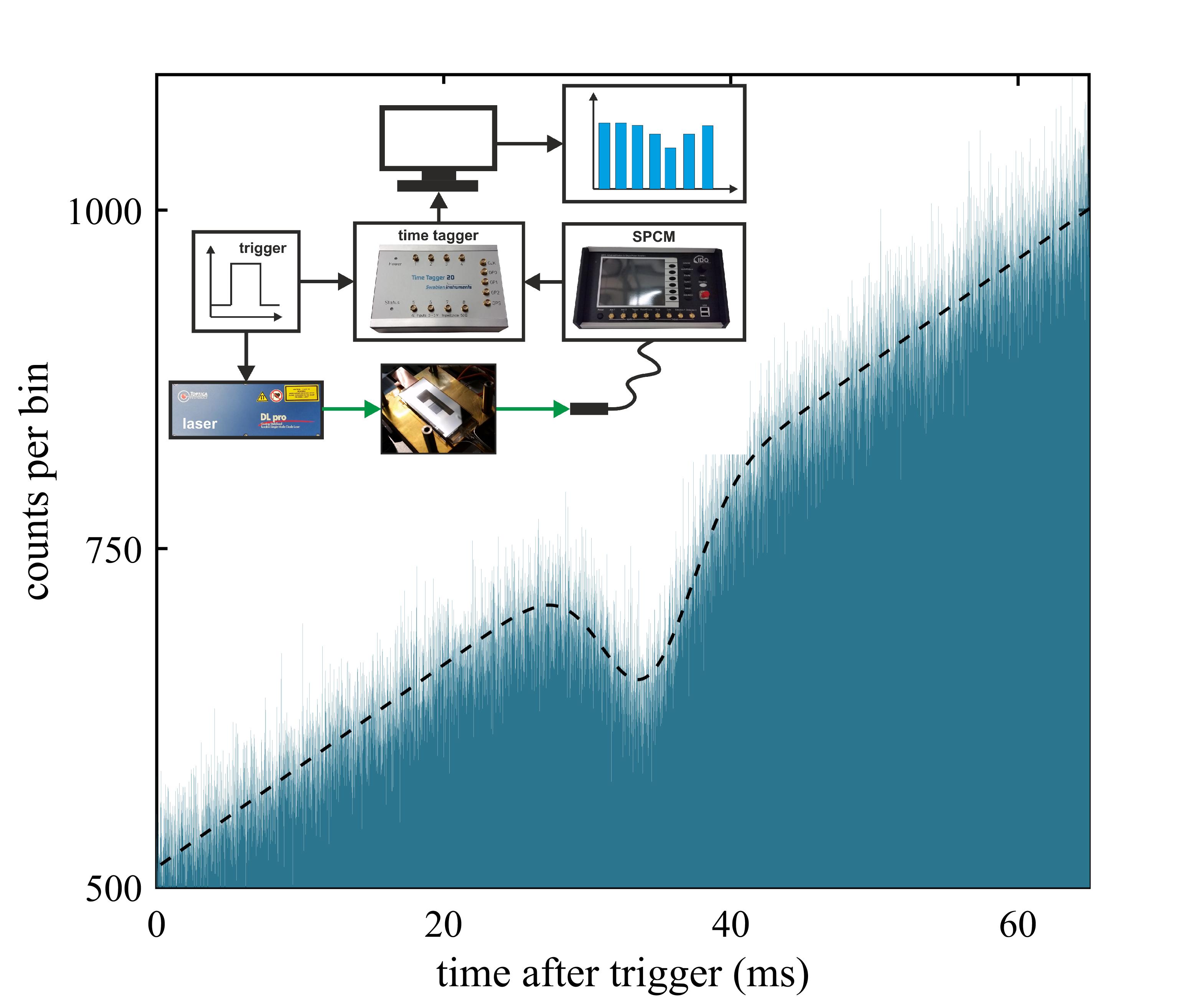}
\caption{\small Histogram of the counted photons transmitted through the nano-device. The black dashed line acts as guide to the eye. The inset sketches the measurement scheme to produce the depicted spectrum.}
\label{fig:Hist}
\end{figure}

The inset of Fig.~\ref{fig:Hist} shows the measurement scheme to obtain atomic spectra with the SPCM. 
A trigger signal sent simultaneously to the laser and the time tagger initiates the counting of photons. 
The laser scans the frequency at a rate of 7.7 Hz (every 130 ms back-and-forth or 65 ms one way). 
Photons traveling through the nano-device are counted by the SPCM.
Those counts are summed every 10 $\mu$s (bin width) in one bin (blue bars in Fig.~\ref{fig:Hist}) after the trigger. 
On atomic resonance, the counts decrease.
The slope in the background is due to changing laser power along the scan. 
The raw data in Fig.~\ref{fig:Hist} is obtained by summing all bins over 2000 consecutive triggers.
In contrast to the traces recorded with the APD combined with a Lock-in amplifier, the SPCM measurements show absolute values for absorption. This allows access to properties like the optical density of the vapor.

\subsection{Impact of Atomic Density}

\begin{figure}[htbp]
\centering
\includegraphics[width=8.21cm]{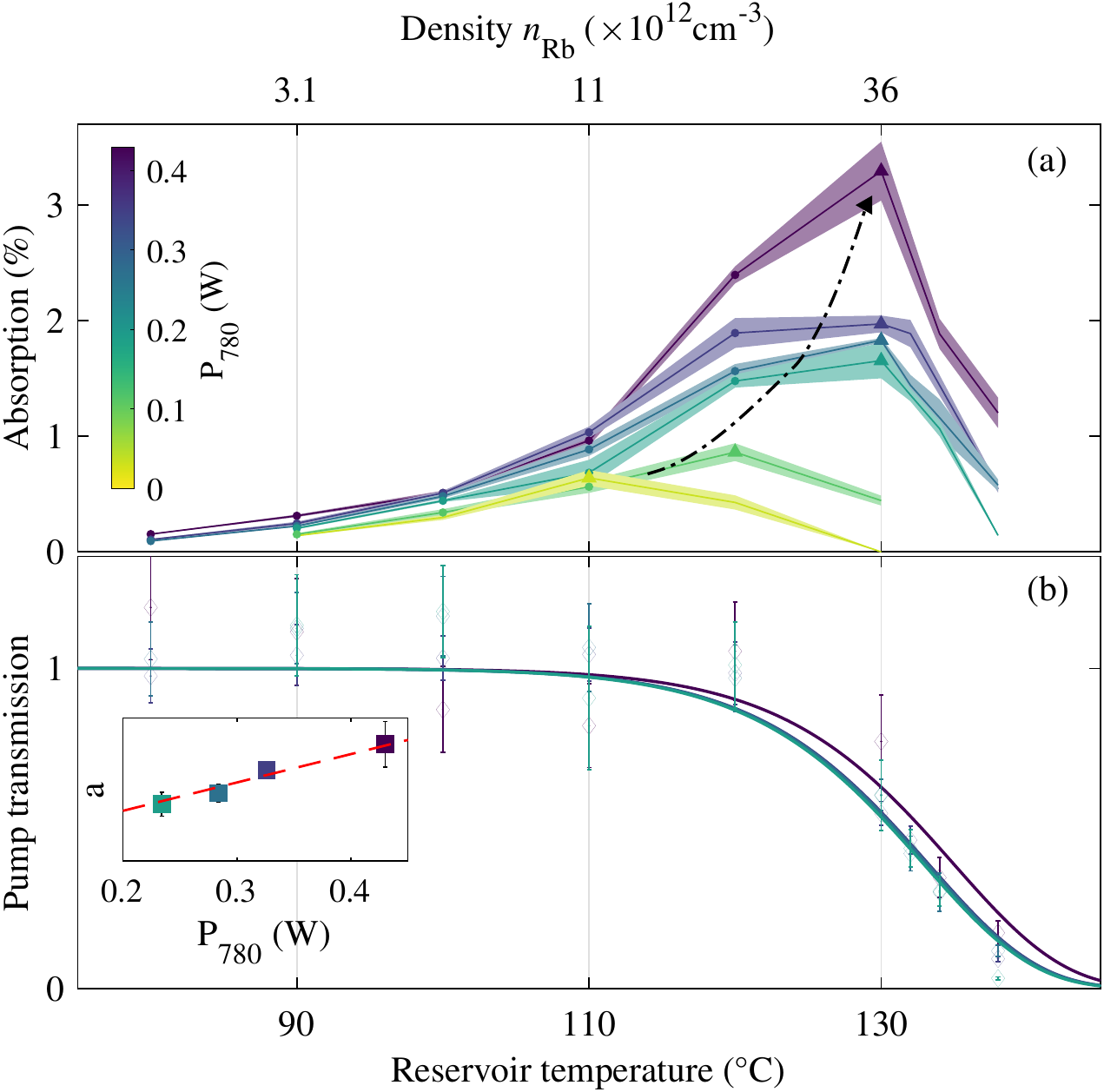}
\caption{\small (a) The measured mean absorption as a function of the ground state density for increasing pump power.
The shaded regions around each curve indicate the standard deviation determined from 4 measurements for each data point.
The traces show a maximum that monotonically increases with density and probe power indicated by the arrow. 
The signal maximum becomes more pronounced with increasing density and shifts towards higher densities. An empirical increase in the density is fitted for each power to the rising slope (circles) in (a).
(b) Density dependent pump transmission for different powers.
Discrete data points (diamonds) and the fit function (solid lines) are color coded similar to the pump power traces in panel (a).
The decrease in absorption at higher densities follows Beer's law which is fitted to the data as solid lines with Eq.~\ref{eq:beer}. 
The inset shows the dependence of density with pump power for the rising slope.
Each colored square corresponds to one fixed pump power with the same color as depicted in panel (a). 
The dashed red line is a guide to the eye showing the linear behavior of this parameter with pump power. 
}
\label{fig:dens}
\end{figure}

The 780 nm pump laser has to travel through a 5 mm thick Rb vapor to reach the chip, whereas the 1529 nm probe laser only interacts via its evanescent tail along a 1 mm long interaction region. 
Hence, the detected two-photon signal is highly dependent on the Rb density within the cell.
The atomic density $n$ influences both the pump and the probe laser. 
To study density effects in this section the reservoir temperature was varied from 80$^\circ$C to 140$^\circ$C, which corresponds to $^{85}$Rb ground state density $n_{\text{Rb}}$ of $\mathcal{O}(10^{12})$ to $\mathcal{O}(10^{13})$ cm$^{-3}$. 
The cell temperature was constantly kept at 160$^\circ$C to prevent Rb condensation on the chip.
All data was recorded with the APD for a fixed probe power of 210 fW out of the waveguide.
Figure~\ref{fig:dens}(a) demonstrates the density on the absorption signal for various pump powers.
For each pump power the shaded region originates from statistical averages over four measurements at each data point. 
For all pump powers at first, the absorption increases with density, controlled by the reservoir temperature until it reaches a maximum. 
This trend can be qualitatively understood as a monotonic increase in the number of atoms, hence more absorption of the probe beam. 
While the effect of atoms on pump propagation can be ignored in dilute regime, at higher densities the 780 nm laser is absorbed significantly.
This leads to a decrease in 5P population and hence a decreased absorption of the probe, consequently.
The absorption maximum moves towards higher densities. 
In order to investigate this behavior more quantitatively, the rising slope (dots on top of each trace) of each curve is fitted with the empirical ground state density expression $f_0 = a(P_{\text{780}})\times n_{\text{Rb}}(T)$. 
The only fit parameter $a$ includes all physical constants and depends on the pump power ($P_{\text{780}}$), since this defines the density in the 5P$_{3/2}$ state.
This parameter $a$ is plotted vs. pump power in the inset of Fig.~\ref{fig:dens}(b). 
The red line is a linear fit and a guide to the eye.
Despite high pump power on a $\approx$ 4 mm$^2$ spot, the signal shows no signature of the optical pumping of the atoms into the dark state manifested by a saturation behavior in the inset of Fig.~\ref{fig:dens}(b). 
All data points are color-coded as the colorbar in Fig.~\ref{fig:dens}. 
Dividing the data in Fig.~\ref{fig:dens}(a) by $f_0$ removes all linear dependence of the data to the atomic density as well as the pump power. 
The rest might be attributed to the absorption of the pump laser which is depicted for the corresponding pump power and density in Fig.~\ref{fig:dens}(b).
Following this approach, the pump transmission $T$ should obey Beer's law.
\begin{equation}
T = T_0\exp(-\alpha d)
\label{eq:beer}
\end{equation}\\
Solid lines represent fits using Eq.~\ref{eq:beer} with $T_0$ as free fit parameter, the absorption coefficient $\alpha$ and propagation length $d$. Errorbars stem from statistical standard deviation from (a) and fit deviations from $f_0$.
The fits show no dependence on pump power since all lines overlap.

\section*{Funding Information}
H. A. acknowledges the support from IQST young researcher award and Eliteprogramm fellowship from Baden-W\"urttemberg foundation. 

\section*{Acknowledgments}
The authors thank R. Ritter, M. Zentile, and D. Weller for several fruitful discussions and technical support. We thank H. Dobbertin and S. Scheel for CP calculations. 
We Also thank M. Kaschel from Institute for Microelectronics Stuttgart (IMS) for helping us with device fabrications.

\section*{Disclosures}
The authors declare no conflicts of interest.
\bibliographystyle{apsrev4-2}
\bibliography{main}

\begin{thebibliography}{52}%
\makeatletter
\providecommand \@ifxundefined [1]{%
 \@ifx{#1\undefined}
}%
\providecommand \@ifnum [1]{%
 \ifnum #1\expandafter \@firstoftwo
 \else \expandafter \@secondoftwo
 \fi
}%
\providecommand \@ifx [1]{%
 \ifx #1\expandafter \@firstoftwo
 \else \expandafter \@secondoftwo
 \fi
}%
\providecommand \natexlab [1]{#1}%
\providecommand \enquote  [1]{``#1''}%
\providecommand \bibnamefont  [1]{#1}%
\providecommand \bibfnamefont [1]{#1}%
\providecommand \citenamefont [1]{#1}%
\providecommand \href@noop [0]{\@secondoftwo}%
\providecommand \href [0]{\begingroup \@sanitize@url \@href}%
\providecommand \@href[1]{\@@startlink{#1}\@@href}%
\providecommand \@@href[1]{\endgroup#1\@@endlink}%
\providecommand \@sanitize@url [0]{\catcode `\\12\catcode `\$12\catcode
  `\&12\catcode `\#12\catcode `\^12\catcode `\_12\catcode `\%12\relax}%
\providecommand \@@startlink[1]{}%
\providecommand \@@endlink[0]{}%
\providecommand \url  [0]{\begingroup\@sanitize@url \@url }%
\providecommand \@url [1]{\endgroup\@href {#1}{\urlprefix }}%
\providecommand \urlprefix  [0]{URL }%
\providecommand \Eprint [0]{\href }%
\providecommand \doibase [0]{https://doi.org/}%
\providecommand \selectlanguage [0]{\@gobble}%
\providecommand \bibinfo  [0]{\@secondoftwo}%
\providecommand \bibfield  [0]{\@secondoftwo}%
\providecommand \translation [1]{[#1]}%
\providecommand \BibitemOpen [0]{}%
\providecommand \bibitemStop [0]{}%
\providecommand \bibitemNoStop [0]{.\EOS\space}%
\providecommand \EOS [0]{\spacefactor3000\relax}%
\providecommand \BibitemShut  [1]{\csname bibitem#1\endcsname}%
\let\auto@bib@innerbib\@empty
\bibitem [{\citenamefont {Politi}\ \emph {et~al.}(2008)\citenamefont {Politi},
  \citenamefont {Cryan}, \citenamefont {Rarity}, \citenamefont {Yu},\ and\
  \citenamefont {O'brien}}]{Politi2008}%
  \BibitemOpen
  \bibfield  {author} {\bibinfo {author} {\bibfnamefont {A.}~\bibnamefont
  {Politi}}, \bibinfo {author} {\bibfnamefont {M.~J.}\ \bibnamefont {Cryan}},
  \bibinfo {author} {\bibfnamefont {J.~G.}\ \bibnamefont {Rarity}}, \bibinfo
  {author} {\bibfnamefont {S.}~\bibnamefont {Yu}},\ and\ \bibinfo {author}
  {\bibfnamefont {J.~L.}\ \bibnamefont {O'brien}},\ }\href
  {https://science.sciencemag.org/content/320/5876/646} {\bibfield  {journal}
  {\bibinfo  {journal} {Science}\ }\textbf {\bibinfo {volume} {320}},\ \bibinfo
  {pages} {646} (\bibinfo {year} {2008})}\BibitemShut {NoStop}%
\bibitem [{\citenamefont {O'brien}\ \emph {et~al.}(2009)\citenamefont
  {O'brien}, \citenamefont {Furusawa},\ and\ \citenamefont
  {Vuckovic}}]{O'brien2009}%
  \BibitemOpen
  \bibfield  {author} {\bibinfo {author} {\bibfnamefont {J.~L.}\ \bibnamefont
  {O'brien}}, \bibinfo {author} {\bibfnamefont {A.}~\bibnamefont {Furusawa}},\
  and\ \bibinfo {author} {\bibfnamefont {J.}~\bibnamefont {Vuckovic}},\ }\href
  {https://www.nature.com/articles/nphoton.2009.229} {\bibfield  {journal}
  {\bibinfo  {journal} {Nature Photonics}\ }\textbf {\bibinfo {volume} {3}},\
  \bibinfo {pages} {687} (\bibinfo {year} {2009})}\BibitemShut {NoStop}%
\bibitem [{\citenamefont {Politi}\ \emph {et~al.}(2009)\citenamefont {Politi},
  \citenamefont {Matthews}, \citenamefont {Thompson},\ and\ \citenamefont
  {O'Brien}}]{Politi2009}%
  \BibitemOpen
  \bibfield  {author} {\bibinfo {author} {\bibfnamefont {A.}~\bibnamefont
  {Politi}}, \bibinfo {author} {\bibfnamefont {J.~C.}\ \bibnamefont
  {Matthews}}, \bibinfo {author} {\bibfnamefont {M.~G.}\ \bibnamefont
  {Thompson}},\ and\ \bibinfo {author} {\bibfnamefont {J.~L.}\ \bibnamefont
  {O'Brien}},\ }\href {https://ieeexplore.ieee.org/abstract/document/5340091}
  {\bibfield  {journal} {\bibinfo  {journal} {IEEE Journal of Selected Topics
  in Quantum Electronics}\ }\textbf {\bibinfo {volume} {15}},\ \bibinfo {pages}
  {1673} (\bibinfo {year} {2009})}\BibitemShut {NoStop}%
\bibitem [{\citenamefont {Dai}\ \emph {et~al.}(2012)\citenamefont {Dai},
  \citenamefont {Bauters},\ and\ \citenamefont {Bowers}}]{Dai2012}%
  \BibitemOpen
  \bibfield  {author} {\bibinfo {author} {\bibfnamefont {D.}~\bibnamefont
  {Dai}}, \bibinfo {author} {\bibfnamefont {J.}~\bibnamefont {Bauters}},\ and\
  \bibinfo {author} {\bibfnamefont {J.~E.}\ \bibnamefont {Bowers}},\ }\href
  {https://www.nature.com/articles/lsa20121} {\bibfield  {journal} {\bibinfo
  {journal} {Light: Science \& Applications}\ }\textbf {\bibinfo {volume}
  {1}},\ \bibinfo {pages} {e1} (\bibinfo {year} {2012})}\BibitemShut {NoStop}%
\bibitem [{\citenamefont {Rickman}(2014)}]{Rickman2014}%
  \BibitemOpen
  \bibfield  {author} {\bibinfo {author} {\bibfnamefont {A.}~\bibnamefont
  {Rickman}},\ }\href {https://www.nature.com/articles/nphoton.2014.175}
  {\bibfield  {journal} {\bibinfo  {journal} {Nature Photonics}\ }\textbf
  {\bibinfo {volume} {8}},\ \bibinfo {pages} {579} (\bibinfo {year}
  {2014})}\BibitemShut {NoStop}%
\bibitem [{\citenamefont {Lin}\ \emph {et~al.}(2017)\citenamefont {Lin},
  \citenamefont {Luo}, \citenamefont {Gu}, \citenamefont {Kimerling},
  \citenamefont {Wada}, \citenamefont {Agarwal},\ and\ \citenamefont
  {Hu}}]{Lin2017}%
  \BibitemOpen
  \bibfield  {author} {\bibinfo {author} {\bibfnamefont {H.}~\bibnamefont
  {Lin}}, \bibinfo {author} {\bibfnamefont {Z.}~\bibnamefont {Luo}}, \bibinfo
  {author} {\bibfnamefont {T.}~\bibnamefont {Gu}}, \bibinfo {author}
  {\bibfnamefont {L.~C.}\ \bibnamefont {Kimerling}}, \bibinfo {author}
  {\bibfnamefont {K.}~\bibnamefont {Wada}}, \bibinfo {author} {\bibfnamefont
  {A.}~\bibnamefont {Agarwal}},\ and\ \bibinfo {author} {\bibfnamefont
  {J.}~\bibnamefont {Hu}},\ }\href
  {https://www.degruyter.com/view/j/nanoph.2018.7.issue-2/nanoph-2017-0085/nanoph-2017-0085.xml}
  {\bibfield  {journal} {\bibinfo  {journal} {Nanophotonics}\ }\textbf
  {\bibinfo {volume} {7}},\ \bibinfo {pages} {0085} (\bibinfo {year}
  {2017})}\BibitemShut {NoStop}%
\bibitem [{\citenamefont {Liang}\ \emph {et~al.}(2009)\citenamefont {Liang},
  \citenamefont {Fiorentino}, \citenamefont {Okumura}, \citenamefont {Chang},
  \citenamefont {Spencer}, \citenamefont {Kuo}, \citenamefont {Fang},
  \citenamefont {Dai}, \citenamefont {Beausoleil},\ and\ \citenamefont
  {Bowers}}]{Liang2009}%
  \BibitemOpen
  \bibfield  {author} {\bibinfo {author} {\bibfnamefont {D.}~\bibnamefont
  {Liang}}, \bibinfo {author} {\bibfnamefont {M.}~\bibnamefont {Fiorentino}},
  \bibinfo {author} {\bibfnamefont {T.}~\bibnamefont {Okumura}}, \bibinfo
  {author} {\bibfnamefont {H.-H.}\ \bibnamefont {Chang}}, \bibinfo {author}
  {\bibfnamefont {D.~T.}\ \bibnamefont {Spencer}}, \bibinfo {author}
  {\bibfnamefont {Y.-H.}\ \bibnamefont {Kuo}}, \bibinfo {author} {\bibfnamefont
  {A.~W.}\ \bibnamefont {Fang}}, \bibinfo {author} {\bibfnamefont
  {D.}~\bibnamefont {Dai}}, \bibinfo {author} {\bibfnamefont {R.~G.}\
  \bibnamefont {Beausoleil}},\ and\ \bibinfo {author} {\bibfnamefont {J.~E.}\
  \bibnamefont {Bowers}},\ }\href
  {https://www.osapublishing.org/oe/abstract.cfm?uri=oe-17-22-20355} {\bibfield
   {journal} {\bibinfo  {journal} {Optics Express}\ }\textbf {\bibinfo {volume}
  {17}},\ \bibinfo {pages} {20355} (\bibinfo {year} {2009})}\BibitemShut
  {NoStop}%
\bibitem [{\citenamefont {Silverstone}\ \emph {et~al.}(2014)\citenamefont
  {Silverstone}, \citenamefont {Bonneau}, \citenamefont {Ohira}, \citenamefont
  {Suzuki}, \citenamefont {Yoshida}, \citenamefont {Iizuka}, \citenamefont
  {Ezaki}, \citenamefont {Natarajan}, \citenamefont {Tanner}, \citenamefont
  {Hadfield}, \citenamefont {Zwiller}, \citenamefont {Marshall}, \citenamefont
  {Rarity}, \citenamefont {O'Brien},\ and\ \citenamefont
  {Thompson}}]{Silverstone2014}%
  \BibitemOpen
  \bibfield  {author} {\bibinfo {author} {\bibfnamefont {J.~W.}\ \bibnamefont
  {Silverstone}}, \bibinfo {author} {\bibfnamefont {D.}~\bibnamefont
  {Bonneau}}, \bibinfo {author} {\bibfnamefont {K.}~\bibnamefont {Ohira}},
  \bibinfo {author} {\bibfnamefont {N.}~\bibnamefont {Suzuki}}, \bibinfo
  {author} {\bibfnamefont {H.}~\bibnamefont {Yoshida}}, \bibinfo {author}
  {\bibfnamefont {N.}~\bibnamefont {Iizuka}}, \bibinfo {author} {\bibfnamefont
  {M.}~\bibnamefont {Ezaki}}, \bibinfo {author} {\bibfnamefont {C.~M.}\
  \bibnamefont {Natarajan}}, \bibinfo {author} {\bibfnamefont {M.~G.}\
  \bibnamefont {Tanner}}, \bibinfo {author} {\bibfnamefont {R.~H.}\
  \bibnamefont {Hadfield}}, \bibinfo {author} {\bibfnamefont {V.}~\bibnamefont
  {Zwiller}}, \bibinfo {author} {\bibfnamefont {G.~D.}\ \bibnamefont
  {Marshall}}, \bibinfo {author} {\bibfnamefont {J.~G.}\ \bibnamefont
  {Rarity}}, \bibinfo {author} {\bibfnamefont {J.~L.}\ \bibnamefont
  {O'Brien}},\ and\ \bibinfo {author} {\bibfnamefont {M.~G.}\ \bibnamefont
  {Thompson}},\ }\href {https://www.nature.com/articles/nphoton.2013.339}
  {\bibfield  {journal} {\bibinfo  {journal} {Nature Photonics}\ }\textbf
  {\bibinfo {volume} {8}},\ \bibinfo {pages} {104} (\bibinfo {year}
  {2014})}\BibitemShut {NoStop}%
\bibitem [{\citenamefont {Zhou}\ \emph {et~al.}(2015)\citenamefont {Zhou},
  \citenamefont {Yin},\ and\ \citenamefont {Michel}}]{Zhou2015}%
  \BibitemOpen
  \bibfield  {author} {\bibinfo {author} {\bibfnamefont {Z.}~\bibnamefont
  {Zhou}}, \bibinfo {author} {\bibfnamefont {B.}~\bibnamefont {Yin}},\ and\
  \bibinfo {author} {\bibfnamefont {J.}~\bibnamefont {Michel}},\ }\href
  {https://www.nature.com/articles/lsa2015131} {\bibfield  {journal} {\bibinfo
  {journal} {Light: Science \& Applications}\ }\textbf {\bibinfo {volume}
  {4}},\ \bibinfo {pages} {e358} (\bibinfo {year} {2015})}\BibitemShut
  {NoStop}%
\bibitem [{\citenamefont {Liu}\ \emph {et~al.}(2009)\citenamefont {Liu},
  \citenamefont {Sun}, \citenamefont {Kimerling},\ and\ \citenamefont
  {Michel}}]{Liu2009}%
  \BibitemOpen
  \bibfield  {author} {\bibinfo {author} {\bibfnamefont {J.}~\bibnamefont
  {Liu}}, \bibinfo {author} {\bibfnamefont {X.}~\bibnamefont {Sun}}, \bibinfo
  {author} {\bibfnamefont {L.~C.}\ \bibnamefont {Kimerling}},\ and\ \bibinfo
  {author} {\bibfnamefont {J.}~\bibnamefont {Michel}},\ }\href
  {https://www.osapublishing.org/ol/abstract.cfm?uri=ol-34-11-1738} {\bibfield
  {journal} {\bibinfo  {journal} {Optics Letters}\ }\textbf {\bibinfo {volume}
  {34}},\ \bibinfo {pages} {1738} (\bibinfo {year} {2009})}\BibitemShut
  {NoStop}%
\bibitem [{\citenamefont {Yin}\ \emph {et~al.}(2012)\citenamefont {Yin},
  \citenamefont {Ning}, \citenamefont {Turkdogan}, \citenamefont {Liu},
  \citenamefont {Nichols},\ and\ \citenamefont {Ning}}]{Yin2012}%
  \BibitemOpen
  \bibfield  {author} {\bibinfo {author} {\bibfnamefont {L.}~\bibnamefont
  {Yin}}, \bibinfo {author} {\bibfnamefont {H.}~\bibnamefont {Ning}}, \bibinfo
  {author} {\bibfnamefont {S.}~\bibnamefont {Turkdogan}}, \bibinfo {author}
  {\bibfnamefont {Z.}~\bibnamefont {Liu}}, \bibinfo {author} {\bibfnamefont
  {P.~L.}\ \bibnamefont {Nichols}},\ and\ \bibinfo {author} {\bibfnamefont
  {C.~Z.}\ \bibnamefont {Ning}},\ }\href
  {https://aip.scitation.org/doi/10.1063/1.4729412} {\bibfield  {journal}
  {\bibinfo  {journal} {Journal of Applied Physics}\ }\textbf {\bibinfo
  {volume} {100}},\ \bibinfo {pages} {241905} (\bibinfo {year}
  {2012})}\BibitemShut {NoStop}%
\bibitem [{\citenamefont {Wang}\ \emph {et~al.}(2013)\citenamefont {Wang},
  \citenamefont {Guo}, \citenamefont {Wang}, \citenamefont {Wang},
  \citenamefont {Yin},\ and\ \citenamefont {Zhou}}]{Wang2013}%
  \BibitemOpen
  \bibfield  {author} {\bibinfo {author} {\bibfnamefont {B.}~\bibnamefont
  {Wang}}, \bibinfo {author} {\bibfnamefont {R.}~\bibnamefont {Guo}}, \bibinfo
  {author} {\bibfnamefont {X.}~\bibnamefont {Wang}}, \bibinfo {author}
  {\bibfnamefont {L.}~\bibnamefont {Wang}}, \bibinfo {author} {\bibfnamefont
  {B.}~\bibnamefont {Yin}},\ and\ \bibinfo {author} {\bibfnamefont
  {Z.}~\bibnamefont {Zhou}},\ }\href
  {https://aip.scitation.org/doi/10.1063/1.4795153} {\bibfield  {journal}
  {\bibinfo  {journal} {Journal of Applied Physics}\ }\textbf {\bibinfo
  {volume} {113}},\ \bibinfo {pages} {103108} (\bibinfo {year}
  {2013})}\BibitemShut {NoStop}%
\bibitem [{\citenamefont {Isshiki}\ \emph {et~al.}(2014)\citenamefont
  {Isshiki}, \citenamefont {Jing}, \citenamefont {Sato}, \citenamefont
  {Nakajima},\ and\ \citenamefont {Kimura}}]{Isshiki2014}%
  \BibitemOpen
  \bibfield  {author} {\bibinfo {author} {\bibfnamefont {H.}~\bibnamefont
  {Isshiki}}, \bibinfo {author} {\bibfnamefont {F.}~\bibnamefont {Jing}},
  \bibinfo {author} {\bibfnamefont {T.}~\bibnamefont {Sato}}, \bibinfo {author}
  {\bibfnamefont {T.}~\bibnamefont {Nakajima}},\ and\ \bibinfo {author}
  {\bibfnamefont {T.}~\bibnamefont {Kimura}},\ }\href
  {https://www.osapublishing.org/prj/abstract.cfm?uri=prj-2-3-A45} {\bibfield
  {journal} {\bibinfo  {journal} {Photonics Research}\ }\textbf {\bibinfo
  {volume} {2}},\ \bibinfo {pages} {2327} (\bibinfo {year} {2014})}\BibitemShut
  {NoStop}%
\bibitem [{\citenamefont {Vivien}\ \emph {et~al.}(2012)\citenamefont {Vivien},
  \citenamefont {Polzer}, \citenamefont {Marris-Morini}, \citenamefont
  {Osmond}, \citenamefont {Hartmann}, \citenamefont {Crozat}, \citenamefont
  {Cassan}, \citenamefont {Kopp}, \citenamefont {Zimmermann},\ and\
  \citenamefont {Fedeli}}]{Vivien2012}%
  \BibitemOpen
  \bibfield  {author} {\bibinfo {author} {\bibfnamefont {L.}~\bibnamefont
  {Vivien}}, \bibinfo {author} {\bibfnamefont {A.}~\bibnamefont {Polzer}},
  \bibinfo {author} {\bibfnamefont {D.}~\bibnamefont {Marris-Morini}}, \bibinfo
  {author} {\bibfnamefont {J.}~\bibnamefont {Osmond}}, \bibinfo {author}
  {\bibfnamefont {J.~M.}\ \bibnamefont {Hartmann}}, \bibinfo {author}
  {\bibfnamefont {P.}~\bibnamefont {Crozat}}, \bibinfo {author} {\bibfnamefont
  {E.}~\bibnamefont {Cassan}}, \bibinfo {author} {\bibfnamefont
  {C.}~\bibnamefont {Kopp}}, \bibinfo {author} {\bibfnamefont {H.}~\bibnamefont
  {Zimmermann}},\ and\ \bibinfo {author} {\bibfnamefont {J.~M.}\ \bibnamefont
  {Fedeli}},\ }\href
  {https://www.osapublishing.org/oe/abstract.cfm?uri=oe-20-2-1096} {\bibfield
  {journal} {\bibinfo  {journal} {Optics Express}\ }\textbf {\bibinfo {volume}
  {20}},\ \bibinfo {pages} {1096} (\bibinfo {year} {2012})}\BibitemShut
  {NoStop}%
\bibitem [{\citenamefont {Zhang}\ \emph {et~al.}(2017)\citenamefont {Zhang},
  \citenamefont {Groote}, \citenamefont {Abbasi}, \citenamefont {Loi},
  \citenamefont {Callaghan}, \citenamefont {Corbett}, \citenamefont {Trindade},
  \citenamefont {Bower},\ and\ \citenamefont {Roelkens}}]{Zhang2017}%
  \BibitemOpen
  \bibfield  {author} {\bibinfo {author} {\bibfnamefont {J.}~\bibnamefont
  {Zhang}}, \bibinfo {author} {\bibfnamefont {A.~D.}\ \bibnamefont {Groote}},
  \bibinfo {author} {\bibfnamefont {A.}~\bibnamefont {Abbasi}}, \bibinfo
  {author} {\bibfnamefont {R.}~\bibnamefont {Loi}}, \bibinfo {author}
  {\bibfnamefont {J.~O.}\ \bibnamefont {Callaghan}}, \bibinfo {author}
  {\bibfnamefont {B.}~\bibnamefont {Corbett}}, \bibinfo {author} {\bibfnamefont
  {A.~J.}\ \bibnamefont {Trindade}}, \bibinfo {author} {\bibfnamefont {C.~A.}\
  \bibnamefont {Bower}},\ and\ \bibinfo {author} {\bibfnamefont
  {G.}~\bibnamefont {Roelkens}},\ }\href
  {https://www.osapublishing.org/oe/abstract.cfm?uri=oe-25-13-14290} {\bibfield
   {journal} {\bibinfo  {journal} {Optics Express}\ }\textbf {\bibinfo {volume}
  {25}},\ \bibinfo {pages} {14290} (\bibinfo {year} {2017})}\BibitemShut
  {NoStop}%
\bibitem [{\citenamefont {Vetsch}\ \emph {et~al.}(2010)\citenamefont {Vetsch},
  \citenamefont {Reitz}, \citenamefont {Sague}, \citenamefont {Schmidt},
  \citenamefont {Dawkins},\ and\ \citenamefont {Rauschenbeutel}}]{Vetsch2010}%
  \BibitemOpen
  \bibfield  {author} {\bibinfo {author} {\bibfnamefont {E.}~\bibnamefont
  {Vetsch}}, \bibinfo {author} {\bibfnamefont {D.}~\bibnamefont {Reitz}},
  \bibinfo {author} {\bibfnamefont {G.}~\bibnamefont {Sague}}, \bibinfo
  {author} {\bibfnamefont {R.}~\bibnamefont {Schmidt}}, \bibinfo {author}
  {\bibfnamefont {S.~T.}\ \bibnamefont {Dawkins}},\ and\ \bibinfo {author}
  {\bibfnamefont {A.}~\bibnamefont {Rauschenbeutel}},\ }\href
  {https://journals.aps.org/prl/abstract/10.1103/PhysRevLett.104.203603}
  {\bibfield  {journal} {\bibinfo  {journal} {Physical Review Letters}\
  }\textbf {\bibinfo {volume} {104}},\ \bibinfo {pages} {203603} (\bibinfo
  {year} {2010})}\BibitemShut {NoStop}%
\bibitem [{\citenamefont {Goban}\ \emph {et~al.}(2012)\citenamefont {Goban},
  \citenamefont {Choi}, \citenamefont {Alton}, \citenamefont {D.~Ding},
  \citenamefont {Pototschnig}, \citenamefont {Thiele}, \citenamefont {Stern},\
  and\ \citenamefont {Kimble}}]{Goban2012}%
  \BibitemOpen
  \bibfield  {author} {\bibinfo {author} {\bibfnamefont {A.}~\bibnamefont
  {Goban}}, \bibinfo {author} {\bibfnamefont {K.~S.}\ \bibnamefont {Choi}},
  \bibinfo {author} {\bibfnamefont {D.~J.}\ \bibnamefont {Alton}}, \bibinfo
  {author} {\bibfnamefont {C.~L.}\ \bibnamefont {D.~Ding}}, \bibinfo {author}
  {\bibfnamefont {M.}~\bibnamefont {Pototschnig}}, \bibinfo {author}
  {\bibfnamefont {T.}~\bibnamefont {Thiele}}, \bibinfo {author} {\bibfnamefont
  {N.~P.}\ \bibnamefont {Stern}},\ and\ \bibinfo {author} {\bibfnamefont
  {H.~J.}\ \bibnamefont {Kimble}},\ }\href
  {https://journals.aps.org/prl/abstract/10.1103/PhysRevLett.109.033603}
  {\bibfield  {journal} {\bibinfo  {journal} {Physical Review Letters}\
  }\textbf {\bibinfo {volume} {109}},\ \bibinfo {pages} {033603} (\bibinfo
  {year} {2012})}\BibitemShut {NoStop}%
\bibitem [{\citenamefont {Mitsch}\ \emph {et~al.}(2014)\citenamefont {Mitsch},
  \citenamefont {Sayrin}, \citenamefont {Albrecht}, \citenamefont
  {Schneeweiss},\ and\ \citenamefont {Rauschenbeutel}}]{Mitsch2014}%
  \BibitemOpen
  \bibfield  {author} {\bibinfo {author} {\bibfnamefont {R.}~\bibnamefont
  {Mitsch}}, \bibinfo {author} {\bibfnamefont {C.}~\bibnamefont {Sayrin}},
  \bibinfo {author} {\bibfnamefont {B.}~\bibnamefont {Albrecht}}, \bibinfo
  {author} {\bibfnamefont {P.}~\bibnamefont {Schneeweiss}},\ and\ \bibinfo
  {author} {\bibfnamefont {A.}~\bibnamefont {Rauschenbeutel}},\ }\href
  {https://www.nature.com/articles/ncomms6713} {\bibfield  {journal} {\bibinfo
  {journal} {Nature Communications}\ }\textbf {\bibinfo {volume} {5}},\
  \bibinfo {pages} {5713} (\bibinfo {year} {2014})}\BibitemShut {NoStop}%
\bibitem [{\citenamefont {Kato}\ and\ \citenamefont {Aoki}(2015)}]{Kato2015}%
  \BibitemOpen
  \bibfield  {author} {\bibinfo {author} {\bibfnamefont {S.}~\bibnamefont
  {Kato}}\ and\ \bibinfo {author} {\bibfnamefont {T.}~\bibnamefont {Aoki}},\
  }\href {https://journals.aps.org/prl/abstract/10.1103/PhysRevLett.115.093603}
  {\bibfield  {journal} {\bibinfo  {journal} {Physical Review Letters}\
  }\textbf {\bibinfo {volume} {115}},\ \bibinfo {pages} {093603} (\bibinfo
  {year} {2015})}\BibitemShut {NoStop}%
\bibitem [{\citenamefont {Corzo}\ \emph {et~al.}(2016)\citenamefont {Corzo},
  \citenamefont {Gouraud}, \citenamefont {Chandra}, \citenamefont {Goban},
  \citenamefont {Sheremet}, \citenamefont {Kupriyanov},\ and\ \citenamefont
  {Laura}}]{Corzo2016}%
  \BibitemOpen
  \bibfield  {author} {\bibinfo {author} {\bibfnamefont {N.~V.}\ \bibnamefont
  {Corzo}}, \bibinfo {author} {\bibfnamefont {B.}~\bibnamefont {Gouraud}},
  \bibinfo {author} {\bibfnamefont {A.}~\bibnamefont {Chandra}}, \bibinfo
  {author} {\bibfnamefont {A.}~\bibnamefont {Goban}}, \bibinfo {author}
  {\bibfnamefont {A.~S.}\ \bibnamefont {Sheremet}}, \bibinfo {author}
  {\bibfnamefont {D.~V.}\ \bibnamefont {Kupriyanov}},\ and\ \bibinfo {author}
  {\bibfnamefont {J.}~\bibnamefont {Laura}},\ }\href
  {https://journals.aps.org/prl/abstract/10.1103/PhysRevLett.117.133603}
  {\bibfield  {journal} {\bibinfo  {journal} {Physical Review Letters}\
  }\textbf {\bibinfo {volume} {117}},\ \bibinfo {pages} {133603} (\bibinfo
  {year} {2016})}\BibitemShut {NoStop}%
\bibitem [{\citenamefont {Kien}\ and\ \citenamefont
  {Rauschenbeutel}(2017)}]{Kien2017}%
  \BibitemOpen
  \bibfield  {author} {\bibinfo {author} {\bibfnamefont {F.~L.}\ \bibnamefont
  {Kien}}\ and\ \bibinfo {author} {\bibfnamefont {A.}~\bibnamefont
  {Rauschenbeutel}},\ }\href
  {https://journals.aps.org/pra/abstract/10.1103/PhysRevA.95.023838} {\bibfield
   {journal} {\bibinfo  {journal} {Physical Review A}\ }\textbf {\bibinfo
  {volume} {95}},\ \bibinfo {pages} {023838} (\bibinfo {year}
  {2017})}\BibitemShut {NoStop}%
\bibitem [{\citenamefont {Thompson}\ \emph {et~al.}(2013)\citenamefont
  {Thompson}, \citenamefont {Tiecke}, \citenamefont {de~Leon}, \citenamefont
  {Feist}, \citenamefont {Akimov}, \citenamefont {Gullans}, \citenamefont
  {Zibrov}, \citenamefont {Vuletic},\ and\ \citenamefont
  {Lukin}}]{Thompson2013}%
  \BibitemOpen
  \bibfield  {author} {\bibinfo {author} {\bibfnamefont {J.~D.}\ \bibnamefont
  {Thompson}}, \bibinfo {author} {\bibfnamefont {T.~G.}\ \bibnamefont
  {Tiecke}}, \bibinfo {author} {\bibfnamefont {N.~P.}\ \bibnamefont {de~Leon}},
  \bibinfo {author} {\bibfnamefont {J.}~\bibnamefont {Feist}}, \bibinfo
  {author} {\bibfnamefont {A.~V.}\ \bibnamefont {Akimov}}, \bibinfo {author}
  {\bibfnamefont {M.}~\bibnamefont {Gullans}}, \bibinfo {author} {\bibfnamefont
  {A.~S.}\ \bibnamefont {Zibrov}}, \bibinfo {author} {\bibfnamefont
  {V.}~\bibnamefont {Vuletic}},\ and\ \bibinfo {author} {\bibfnamefont {M.~D.}\
  \bibnamefont {Lukin}},\ }\href
  {https://science.sciencemag.org/content/340/6137/1202} {\bibfield  {journal}
  {\bibinfo  {journal} {Science}\ }\textbf {\bibinfo {volume} {340}},\ \bibinfo
  {pages} {1202} (\bibinfo {year} {2013})}\BibitemShut {NoStop}%
\bibitem [{\citenamefont {Tiecke}\ \emph {et~al.}(2014)\citenamefont {Tiecke},
  \citenamefont {Thompson}, \citenamefont {de~Leon}, \citenamefont {Liu},
  \citenamefont {Vuletic},\ and\ \citenamefont {Lukin}}]{Tiecke2014}%
  \BibitemOpen
  \bibfield  {author} {\bibinfo {author} {\bibfnamefont {T.~G.}\ \bibnamefont
  {Tiecke}}, \bibinfo {author} {\bibfnamefont {J.~D.}\ \bibnamefont
  {Thompson}}, \bibinfo {author} {\bibfnamefont {N.~P.}\ \bibnamefont
  {de~Leon}}, \bibinfo {author} {\bibfnamefont {L.~R.}\ \bibnamefont {Liu}},
  \bibinfo {author} {\bibfnamefont {V.}~\bibnamefont {Vuletic}},\ and\ \bibinfo
  {author} {\bibfnamefont {M.~D.}\ \bibnamefont {Lukin}},\ }\href
  {https://www.nature.com/articles/nature13188} {\bibfield  {journal} {\bibinfo
   {journal} {Nature}\ }\textbf {\bibinfo {volume} {508}},\ \bibinfo {pages}
  {241} (\bibinfo {year} {2014})}\BibitemShut {NoStop}%
\bibitem [{\citenamefont {Goban}\ \emph {et~al.}(2015)\citenamefont {Goban},
  \citenamefont {Hung}, \citenamefont {Hood}, \citenamefont {Yu}, \citenamefont
  {Muniz}, \citenamefont {Painter},\ and\ \citenamefont {Kimble}}]{Goban2015}%
  \BibitemOpen
  \bibfield  {author} {\bibinfo {author} {\bibfnamefont {A.}~\bibnamefont
  {Goban}}, \bibinfo {author} {\bibfnamefont {C.-L.}\ \bibnamefont {Hung}},
  \bibinfo {author} {\bibfnamefont {J.}~\bibnamefont {Hood}}, \bibinfo {author}
  {\bibfnamefont {S.-P.}\ \bibnamefont {Yu}}, \bibinfo {author} {\bibfnamefont
  {J.}~\bibnamefont {Muniz}}, \bibinfo {author} {\bibfnamefont
  {O.}~\bibnamefont {Painter}},\ and\ \bibinfo {author} {\bibfnamefont
  {H.}~\bibnamefont {Kimble}},\ }\href
  {https://journals.aps.org/prl/abstract/10.1103/PhysRevLett.115.063601}
  {\bibfield  {journal} {\bibinfo  {journal} {Physical Review Letters}\
  }\textbf {\bibinfo {volume} {115}},\ \bibinfo {pages} {063601} (\bibinfo
  {year} {2015})}\BibitemShut {NoStop}%
\bibitem [{\citenamefont {Perez-Rios}\ \emph {et~al.}(2017)\citenamefont
  {Perez-Rios}, \citenamefont {Kim},\ and\ \citenamefont {Hung}}]{Perez2017}%
  \BibitemOpen
  \bibfield  {author} {\bibinfo {author} {\bibfnamefont {J.}~\bibnamefont
  {Perez-Rios}}, \bibinfo {author} {\bibfnamefont {M.~E.}\ \bibnamefont
  {Kim}},\ and\ \bibinfo {author} {\bibfnamefont {C.-L.}\ \bibnamefont
  {Hung}},\ }\href@noop {} {\bibfield  {journal} {\bibinfo  {journal} {New
  Journal of Physics}\ }\textbf {\bibinfo {volume} {19}},\ \bibinfo {pages}
  {123035} (\bibinfo {year} {2017})}\BibitemShut {NoStop}%
\bibitem [{\citenamefont {Kulas}\ \emph {et~al.}(2017)\citenamefont {Kulas},
  \citenamefont {Vogt}, \citenamefont {Resch}, \citenamefont {Hartwig},
  \citenamefont {Ganske}, \citenamefont {Matthias}, \citenamefont {Schlippert},
  \citenamefont {Wendrich}, \citenamefont {Ertmer}, \citenamefont {Rasel},
  \citenamefont {Damjanic}, \citenamefont {We{\ss}els}, \citenamefont
  {Kohfeldt}, \citenamefont {Luvsandamdin}, \citenamefont {Schiemangk},
  \citenamefont {Grzeschik}, \citenamefont {Krutzik}, \citenamefont {Wicht},
  \citenamefont {Peters}, \citenamefont {Herrmann},\ and\ \citenamefont
  {L\"ammerzahl}}]{Kulas2017}%
  \BibitemOpen
  \bibfield  {author} {\bibinfo {author} {\bibfnamefont {S.}~\bibnamefont
  {Kulas}}, \bibinfo {author} {\bibfnamefont {C.}~\bibnamefont {Vogt}},
  \bibinfo {author} {\bibfnamefont {A.}~\bibnamefont {Resch}}, \bibinfo
  {author} {\bibfnamefont {J.}~\bibnamefont {Hartwig}}, \bibinfo {author}
  {\bibfnamefont {S.}~\bibnamefont {Ganske}}, \bibinfo {author} {\bibfnamefont
  {J.}~\bibnamefont {Matthias}}, \bibinfo {author} {\bibfnamefont
  {D.}~\bibnamefont {Schlippert}}, \bibinfo {author} {\bibfnamefont
  {T.}~\bibnamefont {Wendrich}}, \bibinfo {author} {\bibfnamefont
  {W.}~\bibnamefont {Ertmer}}, \bibinfo {author} {\bibfnamefont
  {E.}~\bibnamefont {Rasel}}, \bibinfo {author} {\bibfnamefont
  {M.}~\bibnamefont {Damjanic}}, \bibinfo {author} {\bibfnamefont
  {P.}~\bibnamefont {We{\ss}els}}, \bibinfo {author} {\bibfnamefont
  {A.}~\bibnamefont {Kohfeldt}}, \bibinfo {author} {\bibfnamefont
  {E.}~\bibnamefont {Luvsandamdin}}, \bibinfo {author} {\bibfnamefont
  {M.}~\bibnamefont {Schiemangk}}, \bibinfo {author} {\bibfnamefont
  {C.}~\bibnamefont {Grzeschik}}, \bibinfo {author} {\bibfnamefont
  {M.}~\bibnamefont {Krutzik}}, \bibinfo {author} {\bibfnamefont
  {A.}~\bibnamefont {Wicht}}, \bibinfo {author} {\bibfnamefont
  {A.}~\bibnamefont {Peters}}, \bibinfo {author} {\bibfnamefont
  {S.}~\bibnamefont {Herrmann}},\ and\ \bibinfo {author} {\bibfnamefont
  {C.}~\bibnamefont {L\"ammerzahl}},\ }\href@noop {} {\bibfield  {journal}
  {\bibinfo  {journal} {Microgravity Science and Technology}\ }\textbf
  {\bibinfo {volume} {29}},\ \bibinfo {pages} {37} (\bibinfo {year}
  {2017})}\BibitemShut {NoStop}%
\bibitem [{\citenamefont {Talker}\ \emph {et~al.}(2019)\citenamefont {Talker},
  \citenamefont {Arora}, \citenamefont {Barash}, \citenamefont {Wilkowski},\
  and\ \citenamefont {Levy}}]{Uriel2}%
  \BibitemOpen
  \bibfield  {author} {\bibinfo {author} {\bibfnamefont {E.}~\bibnamefont
  {Talker}}, \bibinfo {author} {\bibfnamefont {P.}~\bibnamefont {Arora}},
  \bibinfo {author} {\bibfnamefont {Y.}~\bibnamefont {Barash}}, \bibinfo
  {author} {\bibfnamefont {D.}~\bibnamefont {Wilkowski}},\ and\ \bibinfo
  {author} {\bibfnamefont {U.}~\bibnamefont {Levy}},\ }\href
  {https://doi.org/10.1364/OE.27.033445} {\bibfield  {journal} {\bibinfo
  {journal} {Opt. Express}\ }\textbf {\bibinfo {volume} {27}},\ \bibinfo
  {pages} {33445} (\bibinfo {year} {2019})}\BibitemShut {NoStop}%
\bibitem [{\citenamefont {Bar-David}\ \emph {et~al.}(2017)\citenamefont
  {Bar-David}, \citenamefont {Stern},\ and\ \citenamefont {Levy}}]{Uriel3}%
  \BibitemOpen
  \bibfield  {author} {\bibinfo {author} {\bibfnamefont {J.}~\bibnamefont
  {Bar-David}}, \bibinfo {author} {\bibfnamefont {L.}~\bibnamefont {Stern}},\
  and\ \bibinfo {author} {\bibfnamefont {U.}~\bibnamefont {Levy}},\ }\href
  {https://doi.org/10.1021/acs.nanolett.6b04740} {\bibfield  {journal}
  {\bibinfo  {journal} {Nano Letters}\ }\textbf {\bibinfo {volume} {17}},\
  \bibinfo {pages} {1127} (\bibinfo {year} {2017})}\BibitemShut {NoStop}%
\bibitem [{\citenamefont {Stern}\ \emph {et~al.}(2013)\citenamefont {Stern},
  \citenamefont {Desiatov}, \citenamefont {Goykhman},\ and\ \citenamefont
  {Levy}}]{Uriel4}%
  \BibitemOpen
  \bibfield  {author} {\bibinfo {author} {\bibfnamefont {L.}~\bibnamefont
  {Stern}}, \bibinfo {author} {\bibfnamefont {B.}~\bibnamefont {Desiatov}},
  \bibinfo {author} {\bibfnamefont {I.}~\bibnamefont {Goykhman}},\ and\
  \bibinfo {author} {\bibfnamefont {U.}~\bibnamefont {Levy}},\ }\href@noop {}
  {\bibfield  {journal} {\bibinfo  {journal} {Nature Communications}\ }\textbf
  {\bibinfo {volume} {4}},\ \bibinfo {pages} {1548} (\bibinfo {year}
  {2013})}\BibitemShut {NoStop}%
\bibitem [{\citenamefont {Stern}\ \emph {et~al.}(2017)\citenamefont {Stern},
  \citenamefont {Desiatov}, \citenamefont {Mazurski},\ and\ \citenamefont
  {Levy}}]{Uriel1}%
  \BibitemOpen
  \bibfield  {author} {\bibinfo {author} {\bibfnamefont {L.}~\bibnamefont
  {Stern}}, \bibinfo {author} {\bibfnamefont {B.}~\bibnamefont {Desiatov}},
  \bibinfo {author} {\bibfnamefont {N.}~\bibnamefont {Mazurski}},\ and\
  \bibinfo {author} {\bibfnamefont {U.}~\bibnamefont {Levy}},\ }\href@noop {}
  {\bibfield  {journal} {\bibinfo  {journal} {Nat. Commun.}\ }\textbf {\bibinfo
  {volume} {8}},\ \bibinfo {pages} {14461} (\bibinfo {year}
  {2017})}\BibitemShut {NoStop}%
\bibitem [{\citenamefont {Yang}\ \emph {et~al.}(2007)\citenamefont {Yang},
  \citenamefont {Conkey}, \citenamefont {Wu}, \citenamefont {Yin},
  \citenamefont {Hawkins},\ and\ \citenamefont {Schmidt}}]{Yang2007}%
  \BibitemOpen
  \bibfield  {author} {\bibinfo {author} {\bibfnamefont {W.}~\bibnamefont
  {Yang}}, \bibinfo {author} {\bibfnamefont {D.~B.}\ \bibnamefont {Conkey}},
  \bibinfo {author} {\bibfnamefont {B.}~\bibnamefont {Wu}}, \bibinfo {author}
  {\bibfnamefont {D.}~\bibnamefont {Yin}}, \bibinfo {author} {\bibfnamefont
  {A.~R.}\ \bibnamefont {Hawkins}},\ and\ \bibinfo {author} {\bibfnamefont
  {H.}~\bibnamefont {Schmidt}},\ }\href@noop {} {\bibfield  {journal} {\bibinfo
   {journal} {Nature Photonics}\ }\textbf {\bibinfo {volume} {1}},\ \bibinfo
  {pages} {331} (\bibinfo {year} {2007})}\BibitemShut {NoStop}%
\bibitem [{\citenamefont {Ritter}\ \emph {et~al.}(2015)\citenamefont {Ritter},
  \citenamefont {Gruhler}, \citenamefont {Pernice}, \citenamefont {K\"ubler},
  \citenamefont {Pfau},\ and\ \citenamefont {L\"ow}}]{Ralf1}%
  \BibitemOpen
  \bibfield  {author} {\bibinfo {author} {\bibfnamefont {R.}~\bibnamefont
  {Ritter}}, \bibinfo {author} {\bibfnamefont {N.}~\bibnamefont {Gruhler}},
  \bibinfo {author} {\bibfnamefont {W.}~\bibnamefont {Pernice}}, \bibinfo
  {author} {\bibfnamefont {H.}~\bibnamefont {K\"ubler}}, \bibinfo {author}
  {\bibfnamefont {T.}~\bibnamefont {Pfau}},\ and\ \bibinfo {author}
  {\bibfnamefont {R.}~\bibnamefont {L\"ow}},\ }\href
  {https://doi.org/10.1063/1.4927172} {\bibfield  {journal} {\bibinfo
  {journal} {Applied Physics Letters}\ }\textbf {\bibinfo {volume} {107}},\
  \bibinfo {pages} {041101} (\bibinfo {year} {2015})}\BibitemShut {NoStop}%
\bibitem [{\citenamefont {Slepkov}\ \emph {et~al.}(2010)\citenamefont
  {Slepkov}, \citenamefont {Bhagwat}, \citenamefont {Venkataraman},
  \citenamefont {Londero}, ,\ and\ \citenamefont {Gaeta}}]{Slepkov2010}%
  \BibitemOpen
  \bibfield  {author} {\bibinfo {author} {\bibfnamefont {A.~D.}\ \bibnamefont
  {Slepkov}}, \bibinfo {author} {\bibfnamefont {A.~R.}\ \bibnamefont
  {Bhagwat}}, \bibinfo {author} {\bibfnamefont {V.}~\bibnamefont
  {Venkataraman}}, \bibinfo {author} {\bibfnamefont {P.}~\bibnamefont
  {Londero}}, ,\ and\ \bibinfo {author} {\bibfnamefont {A.~L.}\ \bibnamefont
  {Gaeta}},\ }\href@noop {} {\bibfield  {journal} {\bibinfo  {journal}
  {Physical Review A}\ }\textbf {\bibinfo {volume} {81}},\ \bibinfo {pages}
  {053825} (\bibinfo {year} {2010})}\BibitemShut {NoStop}%
\bibitem [{\citenamefont {Epple}\ \emph {et~al.}(2014)\citenamefont {Epple},
  \citenamefont {Kleinbach}, \citenamefont {Euser}, \citenamefont {Joly},
  \citenamefont {Pfau}, \citenamefont {Russell},\ and\ \citenamefont
  {L\"ow}}]{Epple2014}%
  \BibitemOpen
  \bibfield  {author} {\bibinfo {author} {\bibfnamefont {G.}~\bibnamefont
  {Epple}}, \bibinfo {author} {\bibfnamefont {K.~S.}\ \bibnamefont
  {Kleinbach}}, \bibinfo {author} {\bibfnamefont {T.~G.}\ \bibnamefont
  {Euser}}, \bibinfo {author} {\bibfnamefont {N.~Y.}\ \bibnamefont {Joly}},
  \bibinfo {author} {\bibfnamefont {T.}~\bibnamefont {Pfau}}, \bibinfo {author}
  {\bibfnamefont {P.~S.~J.}\ \bibnamefont {Russell}},\ and\ \bibinfo {author}
  {\bibfnamefont {R.}~\bibnamefont {L\"ow}},\ }\href@noop {} {\bibfield
  {journal} {\bibinfo  {journal} {Nature Communications}\ }\textbf {\bibinfo
  {volume} {5}},\ \bibinfo {pages} {4132} (\bibinfo {year} {2014})}\BibitemShut
  {NoStop}%
\bibitem [{\citenamefont {Spillane}\ \emph {et~al.}(2008)\citenamefont
  {Spillane}, \citenamefont {Pati}, \citenamefont {Salit}, \citenamefont
  {Hall}, \citenamefont {Kumar}, \citenamefont {Beausoleil}, ,\ and\
  \citenamefont {Shahriar}}]{Spillane2008}%
  \BibitemOpen
  \bibfield  {author} {\bibinfo {author} {\bibfnamefont {S.~M.}\ \bibnamefont
  {Spillane}}, \bibinfo {author} {\bibfnamefont {G.~S.}\ \bibnamefont {Pati}},
  \bibinfo {author} {\bibfnamefont {K.}~\bibnamefont {Salit}}, \bibinfo
  {author} {\bibfnamefont {M.}~\bibnamefont {Hall}}, \bibinfo {author}
  {\bibfnamefont {P.}~\bibnamefont {Kumar}}, \bibinfo {author} {\bibfnamefont
  {R.~G.}\ \bibnamefont {Beausoleil}}, ,\ and\ \bibinfo {author} {\bibfnamefont
  {M.~S.}\ \bibnamefont {Shahriar}},\ }\href@noop {} {\bibfield  {journal}
  {\bibinfo  {journal} {Physical Review Letters}\ }\textbf {\bibinfo {volume}
  {100}},\ \bibinfo {pages} {233602} (\bibinfo {year} {2008})}\BibitemShut
  {NoStop}%
\bibitem [{\citenamefont {Hendrickson}\ \emph {et~al.}(2010)\citenamefont
  {Hendrickson}, \citenamefont {Lai}, \citenamefont {Pittman},\ and\
  \citenamefont {Franson}}]{Hendrickson2010}%
  \BibitemOpen
  \bibfield  {author} {\bibinfo {author} {\bibfnamefont {S.~M.}\ \bibnamefont
  {Hendrickson}}, \bibinfo {author} {\bibfnamefont {M.~M.}\ \bibnamefont
  {Lai}}, \bibinfo {author} {\bibfnamefont {T.~B.}\ \bibnamefont {Pittman}},\
  and\ \bibinfo {author} {\bibfnamefont {J.~D.}\ \bibnamefont {Franson}},\
  }\href@noop {} {\bibfield  {journal} {\bibinfo  {journal} {Physical Review
  Letters}\ }\textbf {\bibinfo {volume} {105}},\ \bibinfo {pages} {173602}
  (\bibinfo {year} {2010})}\BibitemShut {NoStop}%
\bibitem [{\citenamefont {Garcia-Fernandez}\ \emph {et~al.}(2011)\citenamefont
  {Garcia-Fernandez}, \citenamefont {Alt}, \citenamefont {Bruse}, \citenamefont
  {Dan}, \citenamefont {anda O.~Rehband}, \citenamefont {Stiebeiner},
  \citenamefont {Wiedemann}, \citenamefont {Meschede},\ and\ \citenamefont
  {Rauschenbeutel}}]{Fernandez2011}%
  \BibitemOpen
  \bibfield  {author} {\bibinfo {author} {\bibfnamefont {R.}~\bibnamefont
  {Garcia-Fernandez}}, \bibinfo {author} {\bibfnamefont {W.}~\bibnamefont
  {Alt}}, \bibinfo {author} {\bibfnamefont {F.}~\bibnamefont {Bruse}}, \bibinfo
  {author} {\bibfnamefont {C.}~\bibnamefont {Dan}}, \bibinfo {author}
  {\bibfnamefont {K.~K.}\ \bibnamefont {anda O.~Rehband}}, \bibinfo {author}
  {\bibfnamefont {A.}~\bibnamefont {Stiebeiner}}, \bibinfo {author}
  {\bibfnamefont {U.}~\bibnamefont {Wiedemann}}, \bibinfo {author}
  {\bibfnamefont {D.}~\bibnamefont {Meschede}},\ and\ \bibinfo {author}
  {\bibfnamefont {A.}~\bibnamefont {Rauschenbeutel}},\ }\href@noop {}
  {\bibfield  {journal} {\bibinfo  {journal} {Applied Physics B}\ }\textbf
  {\bibinfo {volume} {105}},\ \bibinfo {pages} {3} (\bibinfo {year}
  {2011})}\BibitemShut {NoStop}%
\bibitem [{\citenamefont {Ritter}\ \emph {et~al.}(2016)\citenamefont {Ritter},
  \citenamefont {Gruhler}, \citenamefont {Pernice}, \citenamefont {K\"ubler},
  \citenamefont {Pfau},\ and\ \citenamefont {L\"ow}}]{Ralf2}%
  \BibitemOpen
  \bibfield  {author} {\bibinfo {author} {\bibfnamefont {R.}~\bibnamefont
  {Ritter}}, \bibinfo {author} {\bibfnamefont {N.}~\bibnamefont {Gruhler}},
  \bibinfo {author} {\bibfnamefont {W.~H.~P.}\ \bibnamefont {Pernice}},
  \bibinfo {author} {\bibfnamefont {H.}~\bibnamefont {K\"ubler}}, \bibinfo
  {author} {\bibfnamefont {T.}~\bibnamefont {Pfau}},\ and\ \bibinfo {author}
  {\bibfnamefont {R.}~\bibnamefont {L\"ow}},\ }\href
  {https://doi.org/10.1088/1367-2630/18/10/103031} {\bibfield  {journal}
  {\bibinfo  {journal} {New Journal of Physics}\ }\textbf {\bibinfo {volume}
  {18}},\ \bibinfo {pages} {103031} (\bibinfo {year} {2016})}\BibitemShut
  {NoStop}%
\bibitem [{\citenamefont {Ritter}\ \emph {et~al.}(2018)\citenamefont {Ritter},
  \citenamefont {Gruhler}, \citenamefont {Dobbertin}, \citenamefont {K\"ubler},
  \citenamefont {Scheel}, \citenamefont {Pernice}, \citenamefont {Pfau},\ and\
  \citenamefont {L\"ow}}]{Ralf3}%
  \BibitemOpen
  \bibfield  {author} {\bibinfo {author} {\bibfnamefont {R.}~\bibnamefont
  {Ritter}}, \bibinfo {author} {\bibfnamefont {N.}~\bibnamefont {Gruhler}},
  \bibinfo {author} {\bibfnamefont {H.}~\bibnamefont {Dobbertin}}, \bibinfo
  {author} {\bibfnamefont {H.}~\bibnamefont {K\"ubler}}, \bibinfo {author}
  {\bibfnamefont {S.}~\bibnamefont {Scheel}}, \bibinfo {author} {\bibfnamefont
  {W.}~\bibnamefont {Pernice}}, \bibinfo {author} {\bibfnamefont
  {T.}~\bibnamefont {Pfau}},\ and\ \bibinfo {author} {\bibfnamefont
  {R.}~\bibnamefont {L\"ow}},\ }\href
  {https://doi.org/10.1103/PhysRevX.8.021032} {\bibfield  {journal} {\bibinfo
  {journal} {Phys. Rev. X}\ }\textbf {\bibinfo {volume} {8}},\ \bibinfo {pages}
  {021032} (\bibinfo {year} {2018})}\BibitemShut {NoStop}%
\bibitem [{\citenamefont {Ripka}\ \emph {et~al.}(2018)\citenamefont {Ripka},
  \citenamefont {K\"ubler}, \citenamefont {L\"ow},\ and\ \citenamefont
  {Pfau}}]{Ripka2018}%
  \BibitemOpen
  \bibfield  {author} {\bibinfo {author} {\bibfnamefont {F.}~\bibnamefont
  {Ripka}}, \bibinfo {author} {\bibfnamefont {H.}~\bibnamefont {K\"ubler}},
  \bibinfo {author} {\bibfnamefont {R.}~\bibnamefont {L\"ow}},\ and\ \bibinfo
  {author} {\bibfnamefont {T.}~\bibnamefont {Pfau}},\ }\href@noop {} {\bibfield
   {journal} {\bibinfo  {journal} {Science}\ }\textbf {\bibinfo {volume}
  {362}},\ \bibinfo {pages} {446} (\bibinfo {year} {2018})}\BibitemShut
  {NoStop}%
\bibitem [{\citenamefont {Newman}\ \emph {et~al.}(2019)\citenamefont {Newman},
  \citenamefont {Maurice}, \citenamefont {Drake}, \citenamefont {Stone},
  \citenamefont {Briles}, \citenamefont {Spencer}, \citenamefont {Fredrick},
  \citenamefont {Li}, \citenamefont {Westly}, \citenamefont {Ilic},
  \citenamefont {Shen}, \citenamefont {Suh}, \citenamefont {Yang},
  \citenamefont {Johnson}, \citenamefont {Johnson}, \citenamefont {an~Kerry
  J.~Vahala}, \citenamefont {Srinivasan}, \citenamefont {Diddams},
  \citenamefont {Kitching}, \citenamefont {Papp}, ,\ and\ \citenamefont
  {Hummon}}]{Newman2019}%
  \BibitemOpen
  \bibfield  {author} {\bibinfo {author} {\bibfnamefont {Z.~L.}\ \bibnamefont
  {Newman}}, \bibinfo {author} {\bibfnamefont {V.}~\bibnamefont {Maurice}},
  \bibinfo {author} {\bibfnamefont {T.}~\bibnamefont {Drake}}, \bibinfo
  {author} {\bibfnamefont {J.~R.}\ \bibnamefont {Stone}}, \bibinfo {author}
  {\bibfnamefont {T.~C.}\ \bibnamefont {Briles}}, \bibinfo {author}
  {\bibfnamefont {D.~T.}\ \bibnamefont {Spencer}}, \bibinfo {author}
  {\bibfnamefont {C.}~\bibnamefont {Fredrick}}, \bibinfo {author}
  {\bibfnamefont {Q.}~\bibnamefont {Li}}, \bibinfo {author} {\bibfnamefont
  {D.}~\bibnamefont {Westly}}, \bibinfo {author} {\bibfnamefont {B.~R.}\
  \bibnamefont {Ilic}}, \bibinfo {author} {\bibfnamefont {B.}~\bibnamefont
  {Shen}}, \bibinfo {author} {\bibfnamefont {M.-G.}\ \bibnamefont {Suh}},
  \bibinfo {author} {\bibfnamefont {K.~Y.}\ \bibnamefont {Yang}}, \bibinfo
  {author} {\bibfnamefont {C.}~\bibnamefont {Johnson}}, \bibinfo {author}
  {\bibfnamefont {D.~M.~S.}\ \bibnamefont {Johnson}}, \bibinfo {author}
  {\bibfnamefont {L.~H.}\ \bibnamefont {an~Kerry J.~Vahala}}, \bibinfo {author}
  {\bibfnamefont {K.}~\bibnamefont {Srinivasan}}, \bibinfo {author}
  {\bibfnamefont {S.~A.}\ \bibnamefont {Diddams}}, \bibinfo {author}
  {\bibfnamefont {J.}~\bibnamefont {Kitching}}, \bibinfo {author}
  {\bibfnamefont {S.~B.}\ \bibnamefont {Papp}}, ,\ and\ \bibinfo {author}
  {\bibfnamefont {M.~T.}\ \bibnamefont {Hummon}},\ }\href@noop {} {\bibfield
  {journal} {\bibinfo  {journal} {Optica}\ }\textbf {\bibinfo {volume} {6}},\
  \bibinfo {pages} {680} (\bibinfo {year} {2019})}\BibitemShut {NoStop}%
\bibitem [{\citenamefont {Guo}\ \emph {et~al.}(2019)\citenamefont {Guo},
  \citenamefont {Feng}, \citenamefont {Yang}, \citenamefont {Yu}, \citenamefont
  {Chen}, \citenamefont {Yuan},\ and\ \citenamefont {Zhang}}]{Guo2019}%
  \BibitemOpen
  \bibfield  {author} {\bibinfo {author} {\bibfnamefont {J.}~\bibnamefont
  {Guo}}, \bibinfo {author} {\bibfnamefont {X.}~\bibnamefont {Feng}}, \bibinfo
  {author} {\bibfnamefont {P.}~\bibnamefont {Yang}}, \bibinfo {author}
  {\bibfnamefont {Z.}~\bibnamefont {Yu}}, \bibinfo {author} {\bibfnamefont
  {L.~Q.}\ \bibnamefont {Chen}}, \bibinfo {author} {\bibfnamefont {C.-H.}\
  \bibnamefont {Yuan}},\ and\ \bibinfo {author} {\bibfnamefont
  {W.}~\bibnamefont {Zhang}},\ }\href@noop {} {\bibfield  {journal} {\bibinfo
  {journal} {Nature Communications}\ }\textbf {\bibinfo {volume} {10}},\
  \bibinfo {pages} {148} (\bibinfo {year} {2019})}\BibitemShut {NoStop}%
\bibitem [{\citenamefont {Ducloy}\ and\ \citenamefont {Fichet}(1991)}]{Ducloy}%
  \BibitemOpen
  \bibfield  {author} {\bibinfo {author} {\bibfnamefont {M.}~\bibnamefont
  {Ducloy}}\ and\ \bibinfo {author} {\bibfnamefont {M.}~\bibnamefont
  {Fichet}},\ }\href {https://doi.org/10.1051/jp2:1991160} {\bibfield
  {journal} {\bibinfo  {journal} {Journal de Physique II}\ }\textbf {\bibinfo
  {volume} {1}},\ \bibinfo {pages} {1429} (\bibinfo {year} {1991})}\BibitemShut
  {NoStop}%
\bibitem [{\citenamefont {Alaeian}\ \emph {et~al.}(2019)\citenamefont
  {Alaeian}, \citenamefont {Ritter}, \citenamefont {Basic}, \citenamefont
  {Loew},\ and\ \citenamefont {Pfau}}]{alaeian2019cavity}%
  \BibitemOpen
  \bibfield  {author} {\bibinfo {author} {\bibfnamefont {H.}~\bibnamefont
  {Alaeian}}, \bibinfo {author} {\bibfnamefont {R.}~\bibnamefont {Ritter}},
  \bibinfo {author} {\bibfnamefont {M.}~\bibnamefont {Basic}}, \bibinfo
  {author} {\bibfnamefont {R.}~\bibnamefont {Loew}},\ and\ \bibinfo {author}
  {\bibfnamefont {T.}~\bibnamefont {Pfau}},\ }\href@noop {} {\bibinfo {title}
  {Cavity {QED} based on thermal atoms interacting with a photonic crystal
  cavity: A feasibility study}} (\bibinfo {year} {2019})\BibitemShut {NoStop}%
\bibitem [{\citenamefont {Park}\ \emph {et~al.}(2019)\citenamefont {Park},
  \citenamefont {Kim},\ and\ \citenamefont {Moon}}]{Moon2019}%
  \BibitemOpen
  \bibfield  {author} {\bibinfo {author} {\bibfnamefont {J.}~\bibnamefont
  {Park}}, \bibinfo {author} {\bibfnamefont {H.}~\bibnamefont {Kim}},\ and\
  \bibinfo {author} {\bibfnamefont {H.~S.}\ \bibnamefont {Moon}},\ }\href
  {https://doi.org/10.1103/PhysRevLett.122.143601} {\bibfield  {journal}
  {\bibinfo  {journal} {Phys. Rev. Lett.}\ }\textbf {\bibinfo {volume} {122}},\
  \bibinfo {pages} {143601} (\bibinfo {year} {2019})}\BibitemShut {NoStop}%
\bibitem [{\citenamefont {Schmid}\ \emph {et~al.}(2014)\citenamefont {Schmid},
  \citenamefont {Steinr\"uck},\ and\ \citenamefont {Gottfried}}]{AsymVoigt}%
  \BibitemOpen
  \bibfield  {author} {\bibinfo {author} {\bibfnamefont {M.}~\bibnamefont
  {Schmid}}, \bibinfo {author} {\bibfnamefont {H.-P.}\ \bibnamefont
  {Steinr\"uck}},\ and\ \bibinfo {author} {\bibfnamefont {J.~M.}\ \bibnamefont
  {Gottfried}},\ }\href {https://doi.org/10.1002/sia.5521} {\bibfield
  {journal} {\bibinfo  {journal} {Surface and Interface Analysis}\ }\textbf
  {\bibinfo {volume} {46}},\ \bibinfo {pages} {505} (\bibinfo {year} {2014})},\
  \Eprint
  {https://arxiv.org/abs/https://onlinelibrary.wiley.com/doi/pdf/10.1002/sia.5521}
  {https://onlinelibrary.wiley.com/doi/pdf/10.1002/sia.5521} \BibitemShut
  {NoStop}%
\bibitem [{\citenamefont {McLean}\ \emph {et~al.}(1994)\citenamefont {McLean},
  \citenamefont {Mitchell},\ and\ \citenamefont {Swanston}}]{NumVoigt}%
  \BibitemOpen
  \bibfield  {author} {\bibinfo {author} {\bibfnamefont {A.}~\bibnamefont
  {McLean}}, \bibinfo {author} {\bibfnamefont {C.}~\bibnamefont {Mitchell}},\
  and\ \bibinfo {author} {\bibfnamefont {D.}~\bibnamefont {Swanston}},\ }\href
  {https://doi.org/https://doi.org/10.1016/0368-2048(94)02189-7} {\bibfield
  {journal} {\bibinfo  {journal} {Journal of Electron Spectroscopy and Related
  Phenomena}\ }\textbf {\bibinfo {volume} {69}},\ \bibinfo {pages} {125 }
  (\bibinfo {year} {1994})}\BibitemShut {NoStop}%
\bibitem [{\citenamefont {Bristow}\ \emph {et~al.}(2007)\citenamefont
  {Bristow}, \citenamefont {Rotenberg},\ and\ \citenamefont {van
  Driel}}]{Kerr3}%
  \BibitemOpen
  \bibfield  {author} {\bibinfo {author} {\bibfnamefont {A.~D.}\ \bibnamefont
  {Bristow}}, \bibinfo {author} {\bibfnamefont {N.}~\bibnamefont {Rotenberg}},\
  and\ \bibinfo {author} {\bibfnamefont {H.~M.}\ \bibnamefont {van Driel}},\
  }\href {https://doi.org/10.1063/1.2737359} {\bibfield  {journal} {\bibinfo
  {journal} {Applied Physics Letters}\ }\textbf {\bibinfo {volume} {90}},\
  \bibinfo {pages} {191104} (\bibinfo {year} {2007})}\BibitemShut {NoStop}%
\bibitem [{\citenamefont {Feng}\ \emph {et~al.}(2019)\citenamefont {Feng},
  \citenamefont {Cong}, \citenamefont {Zhang}, \citenamefont {Wei},
  \citenamefont {Liang}, \citenamefont {Fang}, \citenamefont {Wang)},\ and\
  \citenamefont {Jianjun~Zhang1}}]{Feng2019}%
  \BibitemOpen
  \bibfield  {author} {\bibinfo {author} {\bibfnamefont {Q.}~\bibnamefont
  {Feng}}, \bibinfo {author} {\bibfnamefont {H.}~\bibnamefont {Cong}}, \bibinfo
  {author} {\bibfnamefont {B.}~\bibnamefont {Zhang}}, \bibinfo {author}
  {\bibfnamefont {W.}~\bibnamefont {Wei}}, \bibinfo {author} {\bibfnamefont
  {Y.}~\bibnamefont {Liang}}, \bibinfo {author} {\bibfnamefont
  {S.}~\bibnamefont {Fang}}, \bibinfo {author} {\bibfnamefont {T.}~\bibnamefont
  {Wang)}},\ and\ \bibinfo {author} {\bibfnamefont {b.}~\bibnamefont
  {Jianjun~Zhang1}},\ }\href@noop {} {\bibfield  {journal} {\bibinfo  {journal}
  {Applied Physics Letters}\ }\textbf {\bibinfo {volume} {114}} (\bibinfo
  {year} {2019})}\BibitemShut {NoStop}%
\bibitem [{\citenamefont {Gorris-Neveux}\ \emph {et~al.}(1996)\citenamefont
  {Gorris-Neveux}, \citenamefont {Monnot}, \citenamefont {Saltiel},
  \citenamefont {Barb\'e}, \citenamefont {Keller},\ and\ \citenamefont
  {Ducloy}}]{TwoPhotonSRS}%
  \BibitemOpen
  \bibfield  {author} {\bibinfo {author} {\bibfnamefont {M.}~\bibnamefont
  {Gorris-Neveux}}, \bibinfo {author} {\bibfnamefont {P.}~\bibnamefont
  {Monnot}}, \bibinfo {author} {\bibfnamefont {S.}~\bibnamefont {Saltiel}},
  \bibinfo {author} {\bibfnamefont {R.}~\bibnamefont {Barb\'e}}, \bibinfo
  {author} {\bibfnamefont {J.-C.}\ \bibnamefont {Keller}},\ and\ \bibinfo
  {author} {\bibfnamefont {M.}~\bibnamefont {Ducloy}},\ }\href
  {https://doi.org/10.1103/PhysRevA.54.3386} {\bibfield  {journal} {\bibinfo
  {journal} {Phys. Rev. A}\ }\textbf {\bibinfo {volume} {54}},\ \bibinfo
  {pages} {3386} (\bibinfo {year} {1996})}\BibitemShut {NoStop}%
\bibitem [{\citenamefont {Failache}\ \emph {et~al.}(2003)\citenamefont
  {Failache}, \citenamefont {Saltiel}, \citenamefont {Fichet}, \citenamefont
  {Bloch},\ and\ \citenamefont {Ducloy}}]{Failache2003}%
  \BibitemOpen
  \bibfield  {author} {\bibinfo {author} {\bibfnamefont {H.}~\bibnamefont
  {Failache}}, \bibinfo {author} {\bibfnamefont {S.}~\bibnamefont {Saltiel}},
  \bibinfo {author} {\bibfnamefont {M.}~\bibnamefont {Fichet}}, \bibinfo
  {author} {\bibfnamefont {D.}~\bibnamefont {Bloch}},\ and\ \bibinfo {author}
  {\bibfnamefont {M.}~\bibnamefont {Ducloy}},\ }\href
  {https://doi.org/10.1140/epjd/e2003-00098-4} {\bibfield  {journal} {\bibinfo
  {journal} {The European Physical Journal D - Atomic, Molecular, Optical and
  Plasma Physics}\ }\textbf {\bibinfo {volume} {23}},\ \bibinfo {pages} {237}
  (\bibinfo {year} {2003})}\BibitemShut {NoStop}%
\bibitem [{\citenamefont {Hafezi}\ \emph {et~al.}(2013)\citenamefont {Hafezi},
  \citenamefont {Mittal}, \citenamefont {Fan}, \citenamefont {Migdall},\ and\
  \citenamefont {Taylor}}]{Hafezi2013}%
  \BibitemOpen
  \bibfield  {author} {\bibinfo {author} {\bibfnamefont {M.}~\bibnamefont
  {Hafezi}}, \bibinfo {author} {\bibfnamefont {S.}~\bibnamefont {Mittal}},
  \bibinfo {author} {\bibfnamefont {J.}~\bibnamefont {Fan}}, \bibinfo {author}
  {\bibfnamefont {A.}~\bibnamefont {Migdall}},\ and\ \bibinfo {author}
  {\bibfnamefont {J.}~\bibnamefont {Taylor}},\ }\href@noop {} {\bibfield
  {journal} {\bibinfo  {journal} {Nature Photonics}\ }\textbf {\bibinfo
  {volume} {7}},\ \bibinfo {pages} {1001} (\bibinfo {year} {2013})}\BibitemShut
  {NoStop}%
\end{thebibliography}%

\end{document}